\documentclass[
 reprint,
 aps,
pra, longbibliography, groupedaddress,nobibnotes
]{revtex4-1}

\usepackage{eqnarray,amsmath,mathtools,nccmath,amssymb}
\usepackage{resizegather}
\usepackage{float}
\usepackage[utf8]{inputenc}
\usepackage{graphicx}
\usepackage{bm}
\usepackage{adjustbox}
\newcommand{\bra}[1]{\langle #1|}
\newcommand{\ket}[1]{|#1\rangle}

\usepackage{xcolor}

\begin{document}


\title{Significant nonclassical paths with atoms and cavities in the double-slit experiment}

\author{J.O. de Almeida$^{1}$}
\email{jessica.almeida@icfo.eu}
\author{M. Lewenstein$^{1,2}$}
\author{J.Q. Quach$^{3}$}%
\affiliation{$^{1}$ICFO - Institut de Ciencies Fotoniques, Barcelona Institute of Science and Technology, Av. Carl Friedrich Gauss 3, 08860 Castelldefels (Barcelona), Spain\\
$^{2}$ICREA, Pg. Lluís Companys 23, 08010 Barcelona, Spain\\
$^{3}$Institute for Photonics and Advanced Sensing and School of Chemistry and Physics,University of Adelaide, South Australia 5005, Australia}

\date{\today}

\begin{abstract}
In the the double-slit experiment, nonclassical paths are Feynman paths that go through both slits. Prior work with atom cavities as which-way detectors in the double-slit experiment, has shown these paths to be experimentally inaccessible. In this paper, we show how such a setup can indeed detect nonclassical paths with 1\% probability, if one considers a different type of nonclassical path than previously investigated. We also show how this setup can be used to erase and restore the coherence of the nonclassical paths. Finally, we also show how atom cavities may be used to implementa exact measure of Born-rule violation [Quach, Which-way double-slit experiments and Born-rule violation, Phys. Rev. A 95, 042129 (2017)], which up until now has only been a formal construct.

\end{abstract}

\pacs{Valid PACS appear here}
\maketitle

\section{\label{sec:level1}Introduction}

Quantum mechanics is undoubtedly one of the most successful theory of the last century. Recent phenomenological developments has lead to a plethora of applications in high precision and sensing tasks~\cite{wisemanmilburn2009,Degen2017}. However, to continue increasing the sensitivity of precision measurements requires a better understanding of the fundamental aspects of the quantum theory~\cite{Khrennikov2011}. Interference and coherence effects are some of the most useful measures in studying quantum mechanical effects. In this work, we will investigate the contributions of nonclassical paths~\cite{Yabuki1986,Sinha2009,Barnum2014,Kauten2017} in the precise measurement of interference effects. 

The double-slit experiment is the foundation of studies in interference effects~\cite{Feynman1963,Cohen,shankar1994}, as well as revealing the wave-nature of matter~\cite{Jonsson1961,Zeilinger1988,greenberger1988,Zeilinger1999}. Typically the nodes (or anti-nodes) are calculated as the result of the path difference arising out of the distance from the slits to the detection screen. This, however, is only an approximation, as first pointed out by Yabuki~\cite{Yabuki1986}. In the Feynman path integral formulation of quantum mechanics~\cite{Feynman1965}, all possible paths between points contribute to the wave function. The direct paths from the slits to the detection screen are just one set of an infinite number of possible paths. \textit{Higher-order }or \textit{nonclassical} paths include paths which enter both slits before reaching the detector, as shown in Fig.~\ref{fig:paths}. Typically, these nonclassical paths are much less probable than the direct or \textit{classical} paths; nevertheless, it has been shown that in regimes, where the wavelength is large compared to the split-spacing, these nonclassical paths can be significant~\cite{Rengaraj2018}. The nonclassical path contributions to the interference pattern is not uniquely a quantum mechanical effect. Such contributions to the interference pattern arise also out of Maxwell's equations, as shown with finite-difference time-domain simulations~\cite{Taflove,Raedt2012,Sinha2015}.

In the double-slit experiment, the particle nature of matter is revealed if one knows which slit the particle went through~\cite{Wooters1979}. In 1991, Scully \textit{et al.}~\cite{Scully1991} introduced cavities into the slits as a means to mark which slit the particle went through, thereby acting as which-way detectors. They  showed how the setup could implement the delayed-choice quantum erasure experiment using the atomic transition due to atom-light interaction~\cite{Scully1982,Wheeler,Zeilinger,Kim,Wu,Walborn2002}. In this setup energy is necessarily exchanged to reveal which-way information. Of note is an alternative setup proposed by Pavičić which uses monolithic total-internal-reflection resonators to perform efficient which-way detection without energy exchange~\cite{Pavicic1996}.

More recently, de Oliveira \textit{et al.}~\cite{deOliveira2017} showed how the setup proposed by Scully can be used to isolate nonclassical paths. Their work consisted of modeling looped trajectories of Rubidium Rydberg atoms in the double-slit experiment, will the result that the probability of detection of these nonclassical paths were too small to be feasible. 

The Born rule states that if a quantum object is represented by a wave function $\psi(\mathbf{r},t)$, than the probability density of detecting it at position $\mathbf{r}$ and time $t$ is given by the absolute square of the wave function~\cite{Born1926},
\begin{equation}
	P(\mathbf{r},t) = \psi^*(\mathbf{r},t)\psi(\mathbf{r},t) = |\psi(\mathbf{r},t)|^2~.
\label{eq:P}
\end{equation}

Despite being a cornerstone of quantum mechanics, a direct test of the Born rule was not attempted until 2010 by Sinha \textit{et al.}~\cite{Sinha2010}. The test was a measure of the Sorkin parameter~\cite{Sorkin1994}, which quantifies nonpairwise interference, in a triple-slit experiment~\cite{Magana2016}. Since the exponent of the Born rule only allows for pairwise interference, a nonzero Sorkin parameter would suggest violation of the Born rule. Shortly after this experiment, it was pointed out that a nonzero Sorkin parameter would not necessarily indicate Born-rule violation~\cite{Sawant2014}; instead it could be a signature of nonclassical paths. Most recently, Quach~\cite{Quach2017} proposed an alternative parameter, using the double-slit experiment with which-way detectors, as a more accurate measure of Born-rule violation. However, the Quach's parameter up until now, has only been a formal construct.

The goals of this paper are to propose a practical setup to detect nonclassical paths and test the validity of the Born-rule. In the first part of this paper (Secs.~\ref{sec:level2} and ~\ref{sec:level3}), we show how a double-slit experiment with an atom-cavity setup, can detect nonclassical paths with $1\%$ probability. Further, we show the delay-choice quantum erasure in the context of nonclassical paths. In the last part of the paper (Sec.~\ref{sec:level4}), we extend the setup to show its applicability in implementing the Quach's parameter.

\section{\label{sec:level2}Classical and nonclassical paths}

\begin{figure}[thpb]  
   \centering
   \includegraphics[scale=0.8]{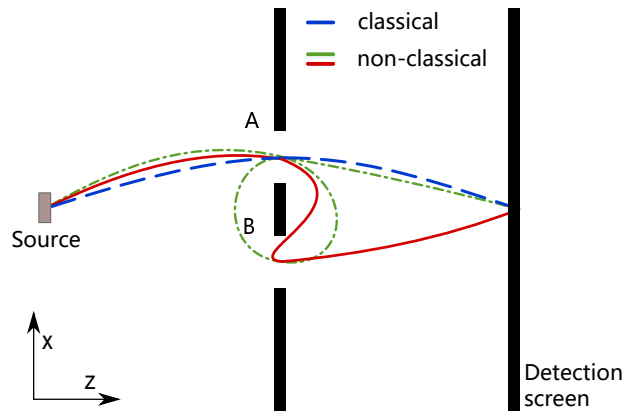}
   \caption{A schematic of the double-slit experiment. The blue-dashed line depicts one the many possible classical paths. The red solid and green dotted-dashed lines depicts two types of nonclassical paths.}
   \label{fig:paths}
\end{figure}

Let us consider the double-slit experiment as depicted in Fig.~\ref{fig:paths}. We make the usual assumption that the slits run infinite in the $y$ direction (perpendicular to the figure plane), and the slit plane extends infinitely in the $x$ direction. This allows us to reduce the system to a one-dimensional problem in the $x$ direction. The source is an atom described by the wave packet
\begin{equation}
\psi_{0}\left(x,t=0\right)=\frac{1}{\sqrt{\sigma_{0} \sqrt{\pi}}} 
\exp{\left[\frac{-x^{2}}{2{\sigma_{0}}^{2}}\right]}~,
\end{equation}
where 
$\sigma_{0}$ is the effective width of the atomic wave packet. The atom wave function at a later time is the weighted sum of all possible paths,  
\begin{equation}
	\psi(x_f,t_f,t_i)=\int_{x_i}K(x_f,t_f;x_i,t_i)\psi_0(x_i,t_i)~,
\end{equation}
where $K(x_f,t_f;x_i,t_i)$ is the free propagator for a particle with mass $m>0$ from point $(x_i,t_i)$ to $(x_f,t_f)$:
\begin{equation} \label{eq:Fpropagator}
K\left(x_{f},t_{f}; x_{i},t_{i} \right)=\sqrt{\frac{m}{2\pi i\hbar\left(t_{f}-t_{i}\right)}} \exp{\left[\frac{i m\left(x_f-x_i\right)^2}{2\hbar\left(t_f-t_i\right)}\right]}~.
\end{equation}

The presence of the slit-plane reduces the number of possible paths between the source and the detection screen. Following the literature we categorize the two types of allowed paths: \textit{classical} paths that go through only one slit, and \textit{nonclassical} paths which go through both slits~\cite{Sawant2014}. Classical and nonclassical paths are also known as \textit{nonexotic} and \textit{exotic}~\cite{deOliveira2017} or \textit{higher-order} paths~\cite{Quach2017}. 

\subsection{\label{sec:clpaths}Classical paths}

The classical paths incorporate all possible paths connecting the source and the detection screen, whenever a single slit, A or B, is open. The wave function resulting from the summation of all paths that go through slit A only  is
\begin{equation}
	\psi_A(x,t)=\int_{x',x_0}K(x,t;x',t')T(x'+d/2)K(x',t';x_0,0)\psi_0(x_0)~,
\label{eq:psi_A}
\end{equation}
where
\begin{equation}
T\left(x\right)=
\exp{\left[\frac{-\left(x\right)^{2}}{2\beta^{2}}\right]}~.
\end{equation}
In Eq.~(\ref{eq:psi_A}), $K(x',t';x_0,0)$ is the free propagator from the source to the slit plane, and $K(x,t;x',t')$ is the free propagator from the slit plane to the detection screen. $T(x')$ is the slit transmission function, which we take to be a Gaussian function of slit-width $\beta$~\cite{deOliveira2017,daPaz2016,Vieira2019}.  
Performing the integral, and taking the limits of integration to infinity yields the following form,
\begin{equation} \label{eq:psiA}
\psi_{A}\left(x\right)= \Gamma_{c} \exp{[\mathrm{c}_{2}x^{2}+\mathrm{c}_{1}x+\mathrm{c}_{0}]}~,
\end{equation}
where the explicit expression for the constants ($\Gamma_{c},\,\mathrm{c}_{2},\,\mathrm{c}_{1},\,\mathrm{c}_{0}$) are given in Appendix~\ref{appendix:ncwavefunction}. The wave function resulting from the summation of classical paths that go through slit $B$ is similarly calculated, 
\begin{equation}
\begin{split}
	\psi_B(x,t)&=\int_{x',x_0}K(x,t;x',t')T(x'-d/2)K(x',t';x_0,0)\psi_0(x_0)\\
    	&=\Gamma_{c} \exp{[\mathrm{c}_{2}x^{2}-\mathrm{c}_{1}x+\mathrm{c}_{0}]}~.
 \end{split}
\label{eq:psiB}
\end{equation}


In the next subsection, we will use this formalism to calculate the nonclassical path that goes through two slits before reaching the detection screen.

\subsection{\label{sec:nclpaths}nonclassical paths}

There are an infinite number of nonclassical paths that enter both slits. nonclassical paths that loop through both slits were considered in the literature~\cite{deOliveira2017,Yabuki1986}. An example of such a path is depicted by the green dotted line in Fig.~\ref{fig:paths}: the particle enters slits A, then slit B, then slit A again, before traveling to the detection screen. In this paper, we will focus on nonclassical paths entering each slit only once. An example of such a path is depicted by the red solid line in Fig.~\ref{fig:paths}: the particle enters slit A, then slit B, then travels to the detection screen. The wave function resulting from the summation of such nonclassical paths is
\begin{equation}\label{eq:psiAB}
\begin{split}
	\psi_{AB}(x,t,\tau)&=\int_{x_{1}',x_{2}',x_{0}} \begin{aligned}[t]&K(x,\Tilde{t}+\tau;x_{2}',\Tilde{t})T(x_{2}'-\frac{d}{2})K(x_{2}',t+\epsilon;x_{1}',t)\\
	&T(x_{1}'+\frac{d}{2})K(x_{1}',t;x_0,0)\psi_{0}(x_{0}) \end{aligned}\\
    	&=\Gamma_{nc} \exp{[\mathrm{c}'_{2}\mathrm{x}^{2}+\mathrm{c}'_{1}x+\mathrm{c}'_{0}]}~,
 \end{split}
\end{equation}
with constants given in Appendix~\ref{appendix:ncwavefunction}. Similarly, the wave function resulting from the summation of nonclassical paths that go through slit $B$ then $A$ is 
\begin{equation}\label{eq:psiBA}
\begin{split}
	\psi_{BA}(x,t,\tau)&=\int_{x_{1}',x_{2}',x_{0}}\begin{aligned}[t]&K(x,\Tilde{t}+\tau;x_{2}',\Tilde{t})T(x_{2}'+\frac{d}{2})K(x_{2}',t+\epsilon;x_{1}',t)\\
	&T(x_{1}'-\frac{d}{2})K(x_{1}',t;x_0,0)\psi_{0}(x_{0})\end{aligned}\\
    	&=\Gamma_{nc} \exp{[\mathrm{c}'_{2}\mathrm{x}^{2}-\mathrm{c}'_{1}x+\mathrm{c}'_{0}]}~,
 \end{split}
\end{equation}
The difference between $\psi_{AB}(x,t,\tau)$ and $\psi_{BA}(x,t,\tau)$ lies in the sign of $\pm c'_1$. 

Using the same formalism, other nonclassical paths can also be calculated. For example, the looped trajectory (green dotted line in Fig.~\ref{fig:paths}), requires an additional transmission through the slits, and therefore has the wave function
\begin{equation}\label{eq:psiBAB}
\begin{split}
	\psi_{BAB}(x,t,\tau)&=\int_{x_{1}',x_{2}',x_{3}',x_{0}}\begin{aligned}[t]&K(x,t'+\tau;x_{3}',t')T(x_{3}'-\frac{d}{2})K(x_{3}',\Tilde{t}+\epsilon;x_{2}',\Tilde{t})T(x_{2}'+\frac{d}{2})\\
	&K(x_{2}',t+\epsilon;x_{1}',t)T(x_{1}'-\frac{d}{2})K(x_{1}',t;x_0,0)\psi_{0}(x_{0})~.\end{aligned}
 \end{split}
\end{equation}

In general, each additional slit transmission attenuates the wave function, such that  $|\psi(x,t)|^2\approx |a^m\psi_0(x,0)|^2$, where $a$ is the attenuation factor and $m$ is the number of times the atom traverses a slit. For classical paths $m=1$, minimal nonclassical paths $m=2$, and single looped paths $m=3$. For the values show in Appendix A, $a\approx 0.1$; as such, the probability of detecting minimal nonclassical paths is 1\% and loop paths is 0.01\%, relative to the classical paths. We will use these facts in the next section to show one may indeed detect minimal nonclassical paths.

\section{\label{sec:level3}Cavity which-way detectors}

\begin{figure}[thpb]  
   \centering
   \includegraphics[scale=0.75]{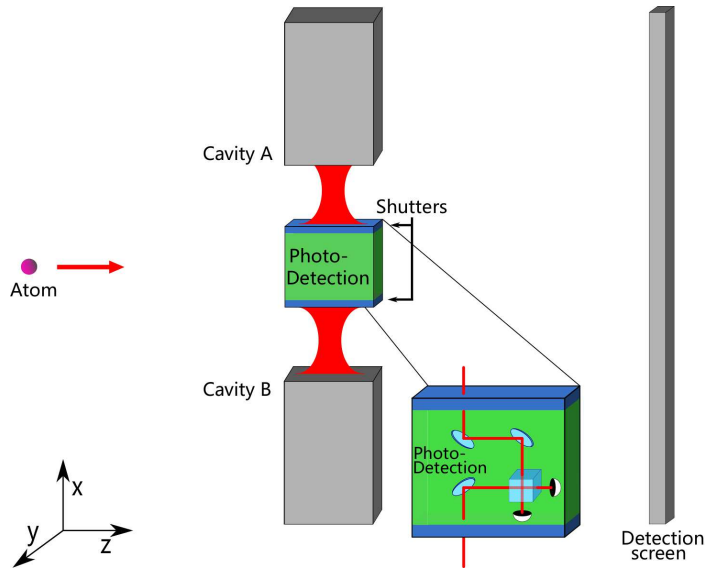}
   \caption{Scheme of atom and double-slit with photonic cavities in each slit. The blue and green box in between the double slit contains the shutters and photodetection scheme. The inset (bottom right) shows a magnified view of the process of photodetection, it describes in detail one possible implementation to detect the cavity photons. In this example, with the opening of the shutters, the cavity photons go through a 50:50 beam splitter before its detection.}
   \label{fig:atomslit}
\end{figure}

Our setup consists of placing a cavity into each of the slits as depicted in Fig.~\ref{fig:atomslit}. The source is a two-level Rydberg atom with ground and excited states $\ket{g}$ or $\ket{e}$. The transition frequency between the two states is resonant with the cavity mode $\Omega$. The initial configuration is such that the atom is in the excited state, and there is one photon in each of the cavities,
\begin{equation}
	\ket{\psi_0} = \ket{e}\ket{1}_A\ket{1}_B~.
\end{equation}

The speed of the atom is tuned so that the interaction time with the cavity is
\begin{equation}
	\tau = \frac{\pi}{\sqrt{n+1}\Omega}~,
\end{equation}
where $n+1$ is the number of excitation in the cavity. This interaction time affects a $\pi$ pulse on the atom~\cite{Krause87}. Here we are interested in the case where $n=1$, to ensure that the transition between $\ket{e}\ket{1}_i$ and $\ket{g}\ket{2}_i$ ($i=A,B$) occurs with unit probability. Therefore our interaction time is set to $\tau = \frac{\pi}{\sqrt{2}\Omega}$.

\subsection{\label{sec:nclpathsdet}Atom-cavity interaction}
Initially the atom is in the excited state and there is a single photon in each cavity. If the atom follows the classical path, it enters a single cavity once; in this case the transition $\ket{e}\ket{1}_i \rightarrow\ket{g}\ket{2}_i$ will occur. If the atom follows the nonclassical path it will enter both cavities. Upon leaving the first cavity the atom emits a photon $\ket{e}\ket{1}_i \rightarrow\ket{g}\ket{2}_i$, and on leaving the second cavity, the atom absorbs a photon $\ket{g}\ket{1}_{i'} \rightarrow\ket{e}\ket{0}_{i'}$ ($i\neq i'$). As such, the system state just before the detection screen is
\begin{equation}  \label{eq:psienc}
\begin{split}
|\psi\rangle=& \frac{1}{\sqrt{N_2}}\left[|g\rangle\left(|2\rangle_{A}|1\rangle_{B}|\psi_{A}\rangle+|1\rangle_{A}|2\rangle_{B}|\psi_{B}\rangle\right)+ \right.\\
&\left. +|e\rangle\left(|2\rangle_{A}|0\rangle_{B}|\psi_{AB}\rangle+|0\rangle_{A}|2\rangle_{B}|\psi_{BA}\rangle\right)\right],
\end{split}
\end{equation}
where $N_i$ is the overall normalization factor, which in general will be dependent on the number of slits ($i$) present. The first term represents the state of the system when the atom's (classical) path traverses though slit-A only: here the atom emits a photon into cavity A. Similarly, the second term represents the state of the system when the atom's (classical) path traverses through slit-B only. The third term represents the state of the system when the atom's (nonclassical) path traverses first through slit-A then slit-B: here the atom emits a photon into cavity A and absorbs a photon in cavity B.  Similarly, the fourth term represents the state of the system when the atom's (nonclassical) path traverses first through slit-B then slit-A. 

Defining the following symmetric and anti-symmetric basis states,
\begin{gather} \label{eq:csym}
|\psi^{\pm}_{c}\rangle=\frac{1}{\sqrt{2}}\left(|\psi_{A}\rangle\pm |\psi_{B}\rangle\right)~, \\
|\psi^{\pm}_{nc}\rangle=\frac{1}{\sqrt{2}}\left(|\psi_{AB}\rangle\pm |\psi_{BA}\rangle\right)~,\\
|\mu^{\pm} \rangle=\frac{1}{\sqrt{2}}\left(|2\rangle_{A}|1\rangle_{B}\pm|1\rangle_{A}|2\rangle_{B}\right)~,\\
|\nu^{\pm} \rangle=\frac{1}{\sqrt{2}}\left(|2\rangle_{A}|0\rangle_{B}\pm|0\rangle_{A}|2\rangle_{B}\right)~,
\end{gather}
we can rewrite the state of the system before the detection screen [Eq.~(\ref {eq:psienc})] as
\begin{equation}\label{eq:psifsadet}
\begin{split}
|\psi'_{f}\rangle=&\frac{1}{\sqrt{N_2}}\left[|g\rangle\left(|\psi^{+}_{c}\rangle|\mu^{+}\rangle + |\psi^{-}_{c}\rangle|\mu^{-}\rangle\right)+\right. \\
&\left.+|e\rangle\left(|\psi^{+}_{nc}\rangle|\nu^{+}\rangle + |\psi^{-}_{nc}\rangle|\nu^{-}\rangle\right)\right].
\end{split}
\end{equation}

Equation~(\ref{eq:psifsadet}) shows that by measuring the state of the atom, we can isolate the classical and nonclassical paths. Keeping count only when an excited atom is detected gives the probability distribution of the nonclassical paths:
\begin{equation}\label{eq:Pncc}
\begin{split}
	P_{e}(x)=&\frac{1}{N_2}\left(|\psi^+_{nc}(x)|^{2}+|\psi^-_{nc}(x)|^{2}\right)\\
	=&\frac{1}{N_2}\left(|\psi_{AB}(x)|^{2}+|\psi_{BA}(x)|^{2}\right)~.
\end{split}
\end{equation}
Conversely, keeping count only when a grounded atom is detected, gives the probability distribution of the classical paths,
\begin{equation}\label{eq:Pcc}
\begin{split}
	P_{g}(x)=&\frac{1}{N_2}\left(|\psi^+_{c}(x)|^{2}+|\psi^-_{c}(x)|^{2}\right)\\
	=&\frac{1}{N_2}\left(|\psi_{A}(x)|^{2}+|\psi_{B}(x)|^{2}\right)~.
\end{split}
\end{equation}

\begin{figure}[thpb]  
   \centering
   \includegraphics[scale=0.9]{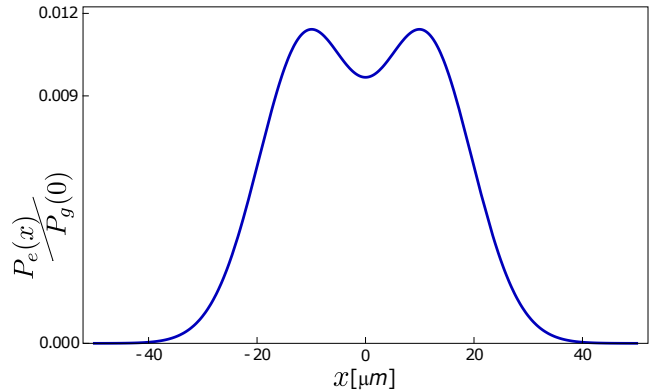}
   \caption{Probability of detecting nonclassical paths normalized by the classical paths. The parameters are defined in Appendix~\ref{appendix:ncwavefunction}.}
   \label{fig:Pnca}
\end{figure}

From Eqs.~(\ref{eq:Pncc}) and (\ref{eq:Pcc}) and the wave functions calculated in previous section, we plot in Fig.~\ref{fig:Pnca} the nonclassical path probability distribution, normalized to the central maximum of the double-slit classical probability distribution, i.e. $P_{e}(x)/P_{g}(0)$.

Figure~\ref{fig:Pnca} shows the spatial distribution of the atoms in the detection screen; the nonclassical paths account for about $1\%$ of the classical paths detection events. The absence of an interference pattern is the result of the distinguishable cavity Fock states that mark the atoms paths. Specifically, from Eq.~(\ref{eq:psienc}) we see that the presence of two photons in cavity A, reveals that the atom went through slit A then B; conversely, when two photons are in cavity B, the atom went through slit B then A. In the next subsection we will show how this which-way information can be erased by introducing shutters and intracavity photodetectors. We will show that erasing this information will retrieve the interference pattern between nonclassical paths.

\subsection{\label{sec:qretriever}Erasing which-way information with cavity photodetection} 

An interesting feature of the atom-cavity implementation of the which-way detectors is that one can partially erase the which-way information and restore coherent interference, even after the atom has been detected. To partially erase the which-way information, we add a beam splitter and photodetectors between the two cavities (Fig.~\ref{fig:atomslit}). Shutters are positioned in each cavity. When the shutters are open, the photons are mixed in a beam splitter device and photodetectors are placed at each output port; the photodetectors act as a reservoir and in the limit of long detection time, all photons present in the cavities are absorbed. This procedure allows to mix the photons from both cavities loosing the which-slit information, retrieving interference. 

The shutters opening and photon detection occurs after the passage of each single atom. The statistics is obtained in the limit of infinite repetitions of this procedure. The beam splitter action on the intracavity photons, corresponds to the following transformation of the $A,B$ input modes:
\begin{equation}\label{eq:a+-}
\hat{a}_{\pm}^\dagger|0\rangle=\frac{\left(a_A\pm a_B\right)^\dagger}{\sqrt{2}}|0\rangle,
\end{equation}
for example in the new basis, the state $|\nu^{-}\rangle=|1 \rangle_{+}|1 \rangle_{-}$. At each output port $+,-$, there is a photodetector, we shall refer to their probability distributions as $P^+$ and $P^-$, respectively. 

We modeled the detection statistics, following the Markovian view of photon absorption~\cite{Srinivas1981}. This strategy predicts photon counts of one, two and three photons in a time interval. The density matrix time evolution contains the detection probabilities at each photon absorption. At time $t=0$ the density matrix is: 
\begin{equation}
\rho(0):=\ket{\psi'_{f}}\bra{\psi'_{f}}~,
\end{equation}
$\ket{\psi'_{f}}$ was defined in Eq.~(\ref{eq:psifsadet}), with photon number in the $+/-$ basis [Eq.~(\ref{eq:a+-})]. The procedure to calculate $\rho(t)$ 
and the probability distributions are described in Appendix~\ref{appendix:rhotprob}~\cite{Zoller1987}. Here we will discuss the most relevant results, and analyze the probabilities in the limits of zero and infinite detection time. 

At zero interaction time, no photons are absorbed, the statistics recover the results of Sec.~\ref{sec:nclpathsdet}. Moreover, for infinite detection time, all photons are absorbed and the number of photons in the cavities is conditioned to the atomic state, as we can see in Eq.~(\ref{eq:psifsadet}). If the atom is detected in the ground state, there are three photons in the cavities, whereas if detected in the excited state, only two photons are in the cavities. 

The probability of measuring the atom in the excited state with two photon counts in the same output photodetector, $P_e^{(kk)}$, or one photon in each output, $P_e^{(kj)}$, $k=+/-,j=+/-$ with $k\neq j$. In the regime of long detection time one gets
\begin{equation}\label{eq:Pesd}
P^{(kk/kj)}_e(x)=\frac{1}{2N_2} \left(|\psi^{(+/-)}_{nc}(x)|^2\right).
\end{equation}
it represents the retrieval of interference, between nonclassical paths $AB$ and $BA$, due to detection of the cavity photons. This is implemented keeping count of the excited atoms, \emph{only} when two photons trigger the same ($P^{(kk)}$) or different ($P^{(kj)}$) detectors.

Similarly, the probability of measuring the atom in the ground state and detecting three photons in the same detector, $P^{(kkk)}$, and two photons in one detector and one in the other, $P^{(kjk)}$ (over all permutations of $kjk$), in the regime of long detection time is:
\begin{equation}\label{eq:Pgst}
P^{(kkk)}_g(x)=\frac{1}{N_2}\left(\frac{3}{4}\right) \left(|\psi^{(k)}_{c}(x)|^2\right),
\end{equation}

\begin{equation}\label{eq:Pgot}
P^{(kjk)}_g(x)=\frac{1}{N_2}\left(\frac{1}{12}\right) \left(|\psi^{(j)}_{c}(x)|^2\right).
\end{equation}
The total probability of atoms in the ground state given by:
\begin{equation}\label{eq:Pgsum}
P_g(x)=\sum_{\substack{k, j=+,-\\ k\neq j}} P_g^{(kkk)}(x)+3P_g^{(kjk)}(x).
\end{equation} 
This result recovers Eq.~(\ref{eq:Pncc}), as expected. This is implemented keeping count of the grounded atoms, \emph{only} when three photons trigger the same or different detectors, respectively.

\begin{figure}[thpb]  
   \centering
   \includegraphics[scale=0.9]{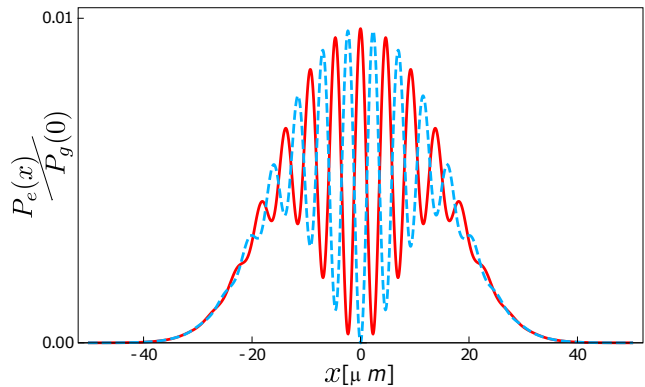}
   \caption{Probability of detecting nonclassical paths in comparison with the classical paths. The red solid curve illustrates the nonclassical paths fringes pattern, it is the total probability of interfering and measuring in a single detector the intra-cavity photons after measuring the atom in the excited state, $P^{(kk)}_{\mathrm{e}}(x)$. The blue dashed curve, shows the anti-fringes pattern, it indicates probability distribution of interfering and measuring the photons in both detectors, $P^{(kj)}_{\mathrm{e}}(x)$ }
   \label{fig:Pnc}
\end{figure}

The magnitude of the nonclassical paths is presented in Fig.~\ref{fig:Pnc}, the interference pattern is recovered, due to the opening of the shutters, which allows for the photons interference and detection. The red solid curve illustrates the nonclassical paths fringes pattern, showing it contributes with up to $1\%$ of the total probability distribution; it is the total probability of interfering and measuring in a single detector the intra-cavity photons, $P^{(kk)}_{\mathrm{e}}(x)$. The blue dashed curve, shows the anti-fringes pattern, it indicates probability distribution of interfering and measuring one photon in each detector, $P^{(kj)}_{e}(x)$.

Intriguingly, at the time at which the atoms were detected, the decision to open or keep closed the photodetector shutter actually had not been made. Whether the atom exhibited coherent interference or not, was determined the time after its detection; we therefore have a manifestation of the delayed-choice quantum erasure experiment. Only when the which-way information is erased by opening the shutters, the atomic inference pattern is retrieved.

In the next subsection, we will discuss how cavities which-way detectors can be used to implement the Quach's parameter and test the Born rule.

\section{\label{sec:level4}The Quach parameter}

The Sorkin parameter for the triple-slit experiment is defined as
\begin{equation}
\mathcal{I}_{ABC} \equiv \mathcal{P}_{ABC}-\mathcal{P}_{AB}-\mathcal{P}_{AC}-\mathcal{P}_{BC}+\mathcal{P}_{A}+\mathcal{P}_{B}+\mathcal{P}_{C}~,
\label{eq:IABC}
\end{equation}
where $\mathcal{P}_{ABC}$ is the probability of detection when all three slits (A,B,C) are open,  $\mathcal{P}_{AB}$ is the probability of detection when two slits (A,B) are open, and so on. If one assumes that the probabilities are simply given by the linear superposition of the individual wave functions of the constituent single-slit setups ($\mathcal{P}_{ABC}=|\psi_A+\psi_B+\psi_C |^2$, $\mathcal{P}_{AB}=|\psi_A+\psi_B |^2$, and so on), then by rewriting probabilities in Eq.~(\ref{eq:IABC}) in terms of wave functions, it can be shown that $\mathcal{I}_{ABC}=0$ if the Born rule is correct. The Sorkin parameter can be generalized to systems with three and more slits, but not two slits. The reason for this is that $\mathcal{I}_{AB}\equiv \mathcal{P}_{AB}-\mathcal{P}_A-\mathcal{P}_B\neq 0$.

If one accounts for nonclassical paths, the probability of detection must be corrected to $\mathcal{P}_{ABC}=|\psi_A+\psi_B+\psi_C+\psi_{ABC} |^2$, where $\psi_{ABC}$ is the wave function made up of nonclassical paths when slits A,B,C are open, which are not accounted for by single-slit wave functions $\psi_A,\psi_B, \psi_C$.  The inclusion of these corrections mean that $I_{ABC}\neq 0$. To overcome these shortcomings, Quach~\cite{Quach2017} proposed an alternative parameter using which-way detectors in a double-slit setup:
\begin{equation}
I_{AB} \equiv \mathcal{P}_{AB}-\mathcal{P}_{D_A}-\mathcal{P}_{D_B}-\mathcal{P}_{D_{AB}}+2\mathcal{P}_{D_AD_B}~,
\label{eq:IAB}
\end{equation}
where

\begin{align}
\mathcal{P}_{D_{A}}(x)&=|\psi_{A}(x)+\psi_{AB}(x)+\psi_{BA}(x)|^{2}+|\psi_{B}(x)|^{2}~,\label{eq:Pda}\\
\mathcal{P}_{D_{B}}(x)&=|\psi_{A}(x)|^{2}+|\psi_{AB}(x)+\psi_{BA}(x)+\psi_{B}(x)|^{2}~,\label{eq:Pdb}\\
\mathcal{P}_{D_{AB}}(x)&=|\psi_{A}(x)+\psi_{B}(x)|^{2}+|\psi_{AB}(x)+\psi_{BA}(x)|^{2}~,\label{eq:Pdab}\\
\mathcal{P}_{D_{A}D_{B}}(x)&=|\psi_{A}(x)|^{2}+|\psi_{B}(x)|^{2}+|\psi_{AB}(x)+\psi_{BA}(x)|^{2}~.\label{eq:Pdadb}
\end{align}

$\mathcal{P}_{D_{A}}(x)$ and $\mathcal{P}_{D_{B}}(x)$ are the probability distributions when there is a which-way detector in slit A or B, respectively. $\mathcal{P}_{D_{A}D_{B}}(x)$ and $\mathcal{P}_{D_{AB}}(x)$ are the probability distributions of distinguishable and indistinguishable which-way detectors in both slits, respectively. Distinguishable which-way detectors identify whether a particle went through slit A or B, indistinguishable which-way detectors knows that a particle went through the slits, but does not know which one. $\psi_{AB}(x)$ consists of Feynman paths that go through slit A first then slit B, and vice versa for $\psi_{BA}(x)$.

Quach's parameter has the advantage that $I_{AB} = 0$ if the Born is rule is not violated, even in the presence of high-order paths and it applies to the double-slit setup. However, up until now Quach's parameter has only been a formal construct. Here we propose how Quach's parameter could be implemented using atom-cavities. 

\subsection{\label{sec:Atom-cavity implementation}Implementation of the Quach parameter}

To implement the Quach parameter we follow the reasoning of Sec.~\ref{sec:level3}, using the cavity as which-way detectors. We also write the parameter in terms of the normalized probability distributions ($P_i$), as this is what is actually measured: 
\begin{equation}
I_{AB} = N_0 P_{AB}-{N_1}\left(P_{D_A}+P_{D_B}\right)-N_2\left(P_{D_{AB}}-2P_{D_AD_B}\right)~,
\label{eq:IABN}
\end{equation}
where $N_i$ are normalization factors that satisfy $\int_{-\infty}^{\infty}\frac{1}{N_0}\mathcal{P}_{AB}(x)dx=$ $\frac{1}{N_1}\int_{-\infty}^{\infty}\mathcal{P}_{D_{A}}(x)dx =$ $\frac{1}{N_2}\int_{-\infty}^{\infty}\mathcal{P}_{D_{AB}}(x)=1$. 

To calculate each of these probability a different initial setup is required, as it is summarized in Table~\ref{table:istates}.

\begin{table}[ht]  
	\begin{center}
		\begin{tabular}{l||c|c|c}  
			\multicolumn{1}{c}{}&\multicolumn{3}{c}{Setup} \\
			\cline{2-4}
			\begin{tabular}{@{}c@{}}Probability \\ distribution\end{tabular}    &      Atom       &    Slit A   &    Slit B     \\
			\hline
			\hline
			$P_{AB}$                    & {}              &  - &  -      \\
			$P_{D_{A}}$      			& $|g\rangle$     &  $|1\rangle$  &  -      \\
			$P_{D_{B}}$      			& {}              &  -  &  $|1\rangle$ \\
			\hline
			\begin{tabular}{@{}c@{}}$P_{D_{AB}}$ \\$P_{D_{A}D_{B}}$  \end{tabular}   	& $|e\rangle$  & $|1\rangle$ & $|1\rangle$  \\
			\hline
			\end{tabular}
		\end{center}
		\caption{Initial setup of the system, to obtain the respective probability distributions. The atom is either intialized in the ground $|g\rangle$ or excited $|e\rangle$ state. $|1\rangle$ represent a slit-cavity initialized with a single photon. }
		\label{table:istates}
\end{table}

To implement the single-slit which-way detector, a cavity is placed in one slit only, leaving the remaining slit empty. To obtain, for example, $P_{D_{A}}(x)$ the setup has a cavity in slit A only, and the atom is initially in the ground state $|g\rangle|1\rangle_{A}|0\rangle_{B}$. The evolved state of the system before the detection screen is
\begin{equation} \label{eq:Psida(b)}
\begin{split}
|\psi_{D_{A}\rangle}=&\frac{1}{\sqrt{N_1}}\left[ |g\rangle|1\rangle_{A}|0\rangle_{B}|\psi_{B}\rangle+\right.\\
&\left. +|e\rangle|0\rangle_{A}|0\rangle_{B}(|\psi_{A}\rangle+|\psi_{AB}\rangle+|\psi_{BA}\rangle)\right].
\end{split}
\end{equation} 
The first term represents the state of the system when the atom traverses slit B \emph{only}. The second term represents the state of the system whenever the atom traversed slit A. In all three cases, the atom absorbed the intracavity photon and transitioned to the excited atomic state $|e\rangle$.

\begin{figure}[tb]
		\includegraphics[width=1.\columnwidth]{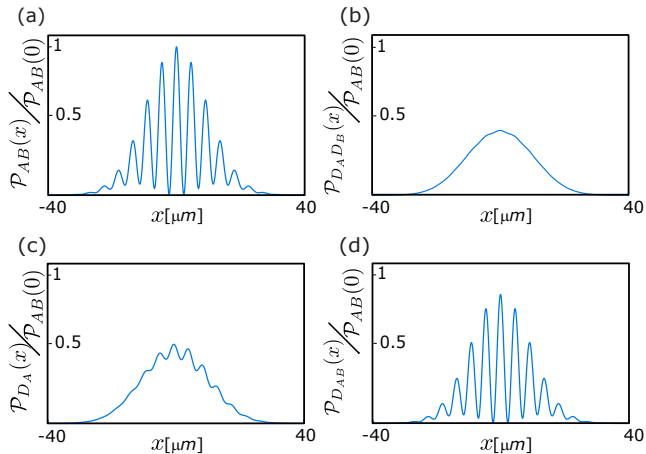}
		\caption{Probability distributions for the different configurations of which-way detectors: a) absence of which-way detectors. b) distinguishable which-way detectors in both slits. c) which-way detector in slit A.  d) indistinguishable which-way detectors in both slits. Summing the probabilities according to Quach's parameter, we obtain exactly zero. }
		\label{fig:Qpar}
\end{figure}

Tracing out the cavity states ($\mathrm{Tr_{c}}$) and projecting on to the position basis, one can retrieve $P_{D_{A}}(x)$:
\begin{equation} \label{eq:Pda(b)}
\begin{split}
P_{D_{A}}(x)=&\langle x|\mathrm{Tr_{c}}\left[\left(|e\rangle\langle e|+|g\rangle\langle g|\right)|\psi_{D_{A}}\rangle\langle \psi_{D_{A}}| \right]|x\rangle\\
=&\frac{1}{N_1}\left(|\psi_{A}(x)+\psi_{AB}(x)+\psi_{BA}(x)|^{2}+|\psi_{B}(x)|^{2}\right).
\end{split}
\end{equation}
In other words, selecting the atoms in the ground and excited state at the detection screen, allows one to obtain the probability of adding a which-way detector in a single slit $P_{D_{A}}(x)$. $P_{D_{B}}(x)$ is similarly obtained with the initial state $|g\rangle|0\rangle_{A}|1\rangle_{B}$. We plot $P_{D_{A}}(x)$ in Fig.~\ref{fig:Qpar}(c) using the wave functions analytically calculated in Sec.~\ref{sec:level2}. $P_{D_{B}}(x)$ has a similar pattern.

To obtain $P_{D_{AB}}(x)$ and $P_{D_{A}D_{B}}(x)$ we use the analysis developed in Sec.~\ref{sec:level3}, i.e. the system is initially $|e\rangle|1\rangle_{A}|1\rangle_{B}$, with the cavity photodetector shutters open. The probability distribution of distinguishable which-way detectors, $P_{D_A D_B}(x)$ is
\begin{equation} \label{eq:Pdadb(b)}
\begin{split}
P_{D_A D_B}(x)=&P_g+2P_e^{(kk)}\\
=&\frac{1}{N_2}\left(|\psi_{A}(x)|^2+|\psi_{B}(x)|^{2}+\right.\\
&\left.+|\psi_{AB}(x)+\psi_{BA}(x)|^{2}\right)~.
\end{split}
\end{equation}
where $P_e^{(kk)}$ is defined in Eq.~(\ref{eq:Pesd}); the factor of $2$ accounts for $k=+$, and $k=-$. $P_g$ is the sum defined in Eq.~(\ref{eq:Pgsum}).

We count all atoms in the ground state, while the ones in the excited state are kept only when two photons trigger the same cavity photodetector. The grounded atoms at each $x$-position give the first term, and the excited atoms give the second term of the probability distribution. We plot $P_{D_A D_B}(x)$ in Fig.~\ref{fig:Qpar}(b).

To calculate the probability distribution of indistinguishable which-way detectors, $P_{D_{AB}}(x)$, we first review Appendix~\ref{appendix:rhotprob}, based on which we can define the required detection strategy. The implementation requires us to keep count of the atoms in the ground state whenever three photons arrive in detector ``$P^+$", and when one photon arrives in ``$P^+$" and two in ``$P^-$". It also requires one to keep count of atoms in the excited state only when two photons reach the same detector,
\begin{equation} \label{eq:Pdab(b)}
\begin{split}
P_{D_{AB}}(x)=&P^{+++}_g+P^{+--}_g+P^{-+-}_g+P^{--+}_g+P_e^{++}+P_e^{--}\\
=&\frac{1}{N_2}\left(\frac{3}{4}|\psi^{+}_{c}(x)|^2+3\left(\frac{1}{12}|\psi^{+}_{c}(x)|^2\right)+|\psi^{+}_{nc}(x)|^2\right)\\
=&\frac{1}{N_2}\left(|\psi_{A}(x)+\psi_{B}(x)|^{2}+|\psi_{AB}(x)+\psi_{BA}(x)|^{2}\right).
\end{split}
\end{equation}
We plot $P_{D_{AB}}(x)$ in Fig.~\ref{fig:Qpar}(d).

With all these probabilities in hand, we calculate $I_{AB}(x)=0$. Obviously, as our theoretical description assumed the Born rule, this result is expected. However, we propose a detailed practical description to test Quach's parameter. In an experiment, $I_{AB}(x)\neq 0$, would implicate a Born-rule violation.

\section{\label{sec:level5}Conclusions}

In this paper, we have shown using Feynman path integrals a class of of nonclassical paths that can be detected with $1\%$ probability. We used an atomic double-slit setup with cavities in each slit to achieve this.  Our proposal operates in the microwave regime, where the nonclassical paths are significant, as the wavelength is large compared to the slits spacing. We also show how our setup may be used to implement a delayed-choice quantum eraser for the nonclassical paths.

As this setup explicitly utilizes quantum mechanical behavior to isolate and detect the nonclassical paths, it offers a possible implementation of Quach's parameter. This is in contrast to other nonclassical path proposals which utilize only the classical nature of light ~\cite{Sawant2014,Sinha2015,Sinha2010}. 


As a future prospect one could use the include the nonclassical paths in studying the Aharonov-Bohm effect~\cite{aharonov1959} in the doubleslit~\cite{Assafrao2019}. Considering the solenoid is strong enough to move the electrons between the slits could be a method to isolate the nonclassical paths.

\section*{ACKNOWLEDGMENTS}
J.O.A and M.L. acknowledges support from ERC AdG NOQIA, Spanish Ministry of Economy and Competitiveness (“Severo Ochoa” program for Centres of Excellence in R\& D (CEX2019-000910-S), Plan National FISICATEAMO and FIDEUA PID2019-106901GB- I00/10.13039/501100011033, FPI), Fundació Privada Cellex, Fundació Mir-Puig, and from Generalitat de Catalunya (AGAUR Grant No. 2017 SGR 1341, CERCA program, QuantumCATU16-011424, co-funded by ERDF Operational Program of Catalonia 2014-2020), MINECO-EU QUANTERA MAQS (funded by State Research Agency (AEI) PCI2019-111828-2/10.13039/501100011033), EU Horizon 2020 FET-OPEN OPTOLogic (Grant No 899794), and the National Science Centre, Poland-Symfonia Grant No. 2016/20/W/ST4/00314. J.Q.Q acknowledges the Ramsay Research Fellowship.

\newpage

\appendix

\section{nonclassical paths wave function}\label{appendix:ncwavefunction}

In this appendix we define the constants used in Eq.( \ref{eq:psiA}-\ref{eq:psiBA}):

\begin{equation}
\begin{split}
\Gamma_{c}=-&\frac{i m \beta}{\pi^{1/4} \sqrt{-i m \sigma_{0} + \frac{t \hbar}{\sigma_{0}}} }. \\
  & \times\frac{1}{\sqrt{\frac{-i m^{2} \beta^{2} \sigma_{0}^{2} + m (t \beta^{2} + (\beta^{2} + \sigma_{0}^{2}) \tau) \hbar + i t \tau \hbar^{2}}{m \sigma_{0}^{2} + i t \hbar}}}
\end{split}
\end{equation}

\begin{equation}
c_{0}= -\frac{m (d^{2} m \sigma_{0}^{2} + i d^{2} (t + \tau) \hbar)}{8 m^{2} \beta^{2} \sigma_{0}^{2} + 8 i m (t \beta^{2} + (\beta^{2} + \sigma_{0}^{2}) \tau) \hbar - 8 t \tau \hbar^{2}}
\end{equation}

\begin{equation}
c_{1}= \frac{m (4 d m \sigma_{0}^{2} + 4 i d t \hbar)}{8 m^{2} \beta^{2} \sigma_{0}^{2} + 8 i m (t \beta^{2} + (\beta^{2} + \sigma_{0}^{2}) \tau) \hbar - 8 t \tau \hbar^{2}}
\end{equation}

\begin{equation}
c_{2}= -\frac{m (4 m (\beta^{2} + \sigma_{0}^{2}) + 4 i t \hbar)}{ 8 m^{2} \beta^{2} \sigma_{0}^{2} + 8 i m (t \beta^{2} + (\beta^{2} + \sigma_{0}^{2}) \tau) \hbar - 8 t \tau \hbar^{2}}
\end{equation}

\begin{widetext}
\begin{minipage}{1\textwidth}
\begin{align}
\Gamma_{nc}=-&\frac{ m^{3/2} \beta^{2} \left(-\frac{1}{\pi}\right)^{1/4}}{\sqrt{-i m\sigma_{0}+\frac{t\hbar}{\sigma_{0}}}}
\frac{1}{\sqrt{\frac{-i m^{2} \beta^{2} \sigma_{0}^{2} + 
  m (t \beta^{2} + \epsilon (\beta^{2} + \sigma_{0}^{2})) \hbar + 
  i t \epsilon \hbar^{2}}{
 m \sigma_{0}^{2} + i t \hbar}}}. \\\nonumber
 &\times\frac{1}{\sqrt{\frac{-i m^{3} \beta^{4} \sigma_{0}^{2} + 
  m^{2} \beta^{2} (t \beta^{2} + \beta^{2} (\epsilon + \tau) + 
\sigma_{0}^{2} (\epsilon + 2 \tau)) \hbar + 
  i m (\epsilon (\beta^{2} + \sigma_{0}^{2}) \tau + 
     t \beta^{2} (\epsilon + 2 \tau)) \hbar^{2} - 
  t \epsilon \tau \hbar^{3}}{
 m^{2} \beta^{2} \sigma_{0}^{2} + 
  i m (t \beta^{2} + \epsilon (\beta^{2} + \sigma_{0}^{2})) \hbar - 
  t \epsilon \hbar^{2}}}}
  \end{align}

\begin{align}
c'_{0}= - \frac{m (2 d^{2} m^{2} \beta^{2} \sigma_{0}^{2} + i d^{2} m (2 t \beta^{2} + 2 \beta^{2} (\epsilon + \tau) + \sigma_{0}^{2} (\epsilon + 4 \tau)) \hbar - d^{2} (\epsilon \tau + t (\epsilon + 4 \tau)) \hbar^{2})}{8 (m^{3} \beta^{4} \sigma_{0}^{2} + 
   i m^{2} \beta^{2} (t \beta^{2} + \beta^{2} (\epsilon + \tau) + \
\sigma_{0}^{2} (\epsilon + 2 \tau)) \hbar - m (\epsilon (\beta^{2} + \sigma_{0}^{2}) \tau + t \beta^{2} (\epsilon + 2 \tau)) \hbar^{2} - i t \epsilon \tau \hbar^{3})}
  \end{align}

\begin{align}
c'_{1}= - \frac{m (4 i d m \epsilon (\beta^{2} + \sigma_{0}^{2}) \hbar - 4 d t \epsilon \hbar^{2})}{8 (m^{3} \beta^{4} \sigma_{0}^{2} + 
   i m^{2} \beta^{2} (t \beta^{2} + \beta^{2} (\epsilon + \tau) + \
\sigma_{0}^{2} (\epsilon + 2 \tau)) \hbar - 
   m (\epsilon (\beta^{2} + \sigma_{0}^{2}) \tau + 
      t \beta^{2} (\epsilon + 2 \tau)) \hbar^{2} - 
   i t \epsilon \tau \hbar^{3})}
  \end{align}

\begin{align}
c'_{2}= - \frac{m (4 m^{2} \beta^{2} (\beta^{2} + 2 \sigma_{0}^{2}) + 4 i m (2 t \beta^{2} + \epsilon (\beta^{2} + \sigma_{0}^{2})) \hbar - 4 t \epsilon \hbar^{2})}{8 (m^{3} \beta^{4} \sigma_{0}^{2} + i m^{2} \beta^{2} (t \beta^{2} + \beta^{2} (\epsilon + \tau) + \sigma_{0}^{2} (\epsilon + 2 \tau)) \hbar - 
   m (\epsilon (\beta^{2} + \sigma_{0}^{2}) \tau + 
      t \beta^{2} (\epsilon + 2 \tau)) \hbar^{2} - 
   i t \epsilon \tau \hbar^{3})}
  \end{align}
\end{minipage}
\end{widetext}

The parameter constants used in this paper are\\
$m=1.44.10^{-25}$kg, $d=5\mu$m, $\sigma_0=\beta=0.3\mu$m, \\ $t=\tau=5$ms, and $\epsilon=2.9$ms.

\newpage

\section{Conditional probability distribution including intra-cavity photodetection}\label{appendix:rhotprob}

The time evolution of the density matrix [state in Eq.~(\ref{eq:psifsadet})], setup with cavities in both slits including the beam splitter and intracavity photodetectors $P^+$ and $P^-$, where the photodetection occurs at individual time intervals, such that the density matrix is calculated solving:

\begin{widetext}
\begin{minipage}{1\textwidth}
\begin{equation}
\begin{split}
\rho(t)&=e^{-\Gamma Nt}\rho(0)e^{-\Gamma Nt} +2\Gamma \sum_{s=-,+}\int_0^t dt' e^{-\Gamma N(t-t')}a_s e^{-\Gamma Nt'} \rho(0) e^{-\Gamma Nt'}a_s^{\dagger}e^{-\Gamma N(t-t')}+\\
 &+(2\Gamma)^2 \sum_{s,s'=-,+}\int_0^t dt' \int_0^{t'} dt'' e^{-\Gamma N(t-t')}a_s e^{-\Gamma N(t'-t'')} a_{s'} e^{-\Gamma Nt''} \rho(0) e^{-\Gamma Nt''} a_{s'}^{\dagger} e^{-\Gamma N(t'-t'')}a_s^{\dagger}e^{-\Gamma N(t-t')}+\\
&+(2\Gamma)^3 \sum_{s,s',s''=-,+}\int_0^t dt' \int_0^{t'} dt''\int_0^{t''} dt''' e^{-\Gamma N(t-t')}a_s e^{-\Gamma N(t'-t'')} a_{s'} e^{-\Gamma N(t''-t''')}a_{s''} e^{-\Gamma N(t''')} \rho(0) e^{-\Gamma Nt'''} a^{\dagger}_{s''} e^{-\Gamma N(t''-t''')} a_{s'}^{\dagger} e^{-\Gamma N(t'-t'')}a_s^{\dagger}e^{-\Gamma N(t-t')}
\label{eq:rhot}
\end{split}
\end{equation}
\end{minipage}
\end{widetext}
where $N$ is the number of photons operator and $\Gamma$ is the cavity width. Equation~(\ref{eq:rhot}) is obtained under the assumption photodetection is a stochastic jump process~\cite{Davies1976,Jacobs2010,wisemanmilburn2009}. 
The first term corresponds to the probability of zero photon absorption, the second term corresponds to the probability of a single photon absorption, the third is the two photon absorption and the last is three photon absorption.

In the limit of long interaction time, $\Gamma t\rightarrow \infty$, both $\frac{1}{6}\left(e^{2\Gamma t}-1\right)^3$ and $\frac{1}{2}\left(e^{2\Gamma t}-1\right)^2$ tend to $1$.

\begin{widetext}
\begin{table}[h]
\begin{minipage}{1\textwidth}
	\begin{center}
	\begin{adjustbox}{width=\textwidth}
		\begin{tabular}{c||c|c|c|}  
			Number of photons & Photodetector & Atom - ground state ($\times\frac{1}{N_2}e^{-6\Gamma t}$) & Atom - excited state ($\times\frac{1}{N_2}e^{-4\Gamma t}$)   \\
			\hline
			\hline
			$0$ & --------- & $\left(|\psi^{+}_{c}|^2+|\psi^{-}_{c}|^2\right)$ & $\left(|\psi^{+}_{nc}|^2+|\psi^{-}_{nc}|^2\right)$\\
			\hline
			$1$ & \begin{tabular}{@{}c@{}}$+$ \\ $-$\end{tabular}  & \begin{tabular}{@{}c@{}} $\left(e^{2\Gamma t}-1\right)\left(\frac{5}{2}|\psi^{+}_{c}|^2+\frac{1}{2}|\psi^{-}_{c}|^2\right)$ \\ $\left(e^{2\Gamma t}-1\right) \left(\frac{1}{2}|\psi^{+}_{c}|^2+\frac{5}{2}|\psi^{-}_{c}|^2\right)$\end{tabular} & \begin{tabular}{@{}c@{}}$\left(e^{2\Gamma t}-1\right)\left(|\psi^{+}_{nc}|^2+|\psi^{-}_{nc}|^2\right)$ \\ $\left(e^{2\Gamma t}-1\right)\left(|\psi^{+}_{nc}|^2+|\psi^{-}_{nc}|^2\right)$\end{tabular}\\
			\hline
			$2$ & \begin{tabular}{@{}c@{}}$++$ \\ $--$ \\ $+-$ or $-+$ \end{tabular}  & \begin{tabular}{@{}c@{}}$\frac{1}{2}\left(e^{2\Gamma t}-1\right)^2 \left(\frac{9}{2}|\psi^{+}_{c}|^2+\frac{1}{2}|\psi^{-}_{c}|^2\right)$ \\ $\frac{1}{2}\left(e^{2\Gamma t}-1\right)^2\left(\frac{1}{2}|\psi^{+}_{c}|^2+\frac{9}{2}|\psi^{-}_{c}|^2\right)$ \\ $\frac{1}{2}\left(e^{2\Gamma t}-1\right)^2\left(\frac{1}{2}|\psi^{+}_{c}|^2+\frac{1}{2}|\psi^{-}_{c}|^2\right)$ \end{tabular} & \begin{tabular}{@{}c@{}}$\frac{1}{2}\left(e^{2\Gamma t}-1\right)^2 \left(|\psi^{+}_{nc}|^2\right)$ \\ $\frac{1}{2}\left(e^{2\Gamma t}-1\right)^2 \left(|\psi^{+}_{nc}|^2\right)$ \\ $\frac{1}{2}\left(e^{2\Gamma t}-1\right)^2\left(|\psi^{-}_{nc}|^2\right)$ \end{tabular} \\
			\hline
			$3$ & \begin{tabular}{@{}c@{}}$+++$ \\ $---$ \\ $+--$ or $-+-$ or $--+$ \\ $++-$ or $-++$ or $+-+$ \end{tabular}  & \begin{tabular}{@{}c@{}}$\frac{1}{6}\left(e^{2\Gamma t}-1\right)^3 \left(\frac{9}{2}|\psi^{+}_{c}|^2\right)$ \\ $\frac{1}{6}\left(e^{2\Gamma t}-1\right)^3 \left(\frac{9}{2}|\psi^{-}_{c}|^2\right)$ \\ $\frac{1}{6}\left(e^{2\Gamma t}-1\right)^3 \left(\frac{1}{2}|\psi^{+}_{c}|^2\right)$ \\ $\frac{1}{6}\left(e^{2\Gamma t}-1\right)^3 \left(\frac{1}{2}|\psi^{-}_{c}|^2\right)$ \end{tabular} &  ---------  \\
			\hline
			\end{tabular}
			\end{adjustbox}
		\end{center}
	
		\caption{Probability distribution at each single photodetection, the order of $+$ and $-$ signs correspond to the temporal order of the photodetection.}
	\label{table:detprob}
\end{minipage}
\end{table}
\end{widetext}

   \bibliography{main}

\begin{thebibliography}{44}%
\makeatletter
\providecommand \@ifxundefined [1]{%
 \@ifx{#1\undefined}
}%
\providecommand \@ifnum [1]{%
 \ifnum #1\expandafter \@firstoftwo
 \else \expandafter \@secondoftwo
 \fi
}%
\providecommand \@ifx [1]{%
 \ifx #1\expandafter \@firstoftwo
 \else \expandafter \@secondoftwo
 \fi
}%
\providecommand \natexlab [1]{#1}%
\providecommand \enquote  [1]{``#1''}%
\providecommand \bibnamefont  [1]{#1}%
\providecommand \bibfnamefont [1]{#1}%
\providecommand \citenamefont [1]{#1}%
\providecommand \href@noop [0]{\@secondoftwo}%
\providecommand \href [0]{\begingroup \@sanitize@url \@href}%
\providecommand \@href[1]{\@@startlink{#1}\@@href}%
\providecommand \@@href[1]{\endgroup#1\@@endlink}%
\providecommand \@sanitize@url [0]{\catcode `\\12\catcode `\$12\catcode
  `\&12\catcode `\#12\catcode `\^12\catcode `\_12\catcode `\%12\relax}%
\providecommand \@@startlink[1]{}%
\providecommand \@@endlink[0]{}%
\providecommand \url  [0]{\begingroup\@sanitize@url \@url }%
\providecommand \@url [1]{\endgroup\@href {#1}{\urlprefix }}%
\providecommand \urlprefix  [0]{URL }%
\providecommand \Eprint [0]{\href }%
\providecommand \doibase [0]{http://dx.doi.org/}%
\providecommand \selectlanguage [0]{\@gobble}%
\providecommand \bibinfo  [0]{\@secondoftwo}%
\providecommand \bibfield  [0]{\@secondoftwo}%
\providecommand \translation [1]{[#1]}%
\providecommand \BibitemOpen [0]{}%
\providecommand \bibitemStop [0]{}%
\providecommand \bibitemNoStop [0]{.\EOS\space}%
\providecommand \EOS [0]{\spacefactor3000\relax}%
\providecommand \BibitemShut  [1]{\csname bibitem#1\endcsname}%
\let\auto@bib@innerbib\@empty
\bibitem [{\citenamefont {Wiseman}\ and\ \citenamefont
  {Milburn}(2009)}]{wisemanmilburn2009}%
  \BibitemOpen
  \bibfield  {author} {\bibinfo {author} {\bibfnamefont {Howard~M.}\
  \bibnamefont {Wiseman}}\ and\ \bibinfo {author} {\bibfnamefont {Gerard~J.}\
  \bibnamefont {Milburn}},\ }\href {\doibase 10.1017/CBO9780511813948} {\emph
  {\bibinfo {title} {Quantum Measurement and Control}}}\ (\bibinfo  {publisher}
  {Cambridge University Press},\ \bibinfo {year} {2009})\BibitemShut {NoStop}%
\bibitem [{\citenamefont {Degen}\ \emph {et~al.}(2017)\citenamefont {Degen},
  \citenamefont {Reinhard},\ and\ \citenamefont {Cappellaro}}]{Degen2017}%
  \BibitemOpen
  \bibfield  {author} {\bibinfo {author} {\bibfnamefont {C.~L.}\ \bibnamefont
  {Degen}}, \bibinfo {author} {\bibfnamefont {F.}~\bibnamefont {Reinhard}}, \
  and\ \bibinfo {author} {\bibfnamefont {P.}~\bibnamefont {Cappellaro}},\
  }\bibfield  {title} {\enquote {\bibinfo {title} {{Quantum sensing}},}\
  }\href@noop {} {\bibfield  {journal} {\bibinfo  {journal} {Rev. Mod. Phys.}\
  }\textbf {\bibinfo {volume} {89}},\ \bibinfo {pages} {1--41} (\bibinfo {year}
  {2017})}\BibitemShut {NoStop}%
\bibitem [{\citenamefont {Khrennikov}(2011)}]{Khrennikov2011}%
  \BibitemOpen
  \bibfield  {author} {\bibinfo {author} {\bibfnamefont {Andrei}\ \bibnamefont
  {Khrennikov}},\ }\bibfield  {title} {\enquote {\bibinfo {title} {Towards
  violation of born’s rule: description of a simple experiment},}\ }\href
  {\doibase 10.1063/1.3567464} {\bibfield  {journal} {\bibinfo  {journal} {AIP
  Conference Proceedings}\ }\textbf {\bibinfo {volume} {1327}},\ \bibinfo
  {pages} {387--393} (\bibinfo {year} {2011})}\BibitemShut {NoStop}%
\bibitem [{\citenamefont {Yabuki}(1986)}]{Yabuki1986}%
  \BibitemOpen
  \bibfield  {author} {\bibinfo {author} {\bibfnamefont {H.}~\bibnamefont
  {Yabuki}},\ }\bibfield  {title} {\enquote {\bibinfo {title} {Feynman path
  integrals in the young double-slit experiment},}\ }\href {\doibase
  10.1007/BF00677704} {\bibfield  {journal} {\bibinfo  {journal} {International
  Journal of Theoretical Physics}\ }\textbf {\bibinfo {volume} {25}},\ \bibinfo
  {pages} {159--174} (\bibinfo {year} {1986})}\BibitemShut {NoStop}%
\bibitem [{\citenamefont {Sinha}\ \emph {et~al.}(2009)\citenamefont {Sinha},
  \citenamefont {Couteau}, \citenamefont {Medendorp}, \citenamefont {Söllner},
  \citenamefont {Laflamme}, \citenamefont {Sorkin},\ and\ \citenamefont
  {Weihs}}]{Sinha2009}%
  \BibitemOpen
  \bibfield  {author} {\bibinfo {author} {\bibfnamefont {Urbasi}\ \bibnamefont
  {Sinha}}, \bibinfo {author} {\bibfnamefont {Christophe}\ \bibnamefont
  {Couteau}}, \bibinfo {author} {\bibfnamefont {Zachari}\ \bibnamefont
  {Medendorp}}, \bibinfo {author} {\bibfnamefont {Immo}\ \bibnamefont
  {Söllner}}, \bibinfo {author} {\bibfnamefont {Raymond}\ \bibnamefont
  {Laflamme}}, \bibinfo {author} {\bibfnamefont {Rafael}\ \bibnamefont
  {Sorkin}}, \ and\ \bibinfo {author} {\bibfnamefont {Gregor}\ \bibnamefont
  {Weihs}},\ }\bibfield  {title} {\enquote {\bibinfo {title} {Testing born’s
  rule in quantum mechanics with a triple slit experiment},}\ }\href {\doibase
  10.1063/1.3109942} {\bibfield  {journal} {\bibinfo  {journal} {AIP Conference
  Proceedings}\ }\textbf {\bibinfo {volume} {1101}},\ \bibinfo {pages}
  {200--207} (\bibinfo {year} {2009})},\ \Eprint
  {http://arxiv.org/abs/https://aip.scitation.org/doi/pdf/10.1063/1.3109942}
  {https://aip.scitation.org/doi/pdf/10.1063/1.3109942} \BibitemShut {NoStop}%
\bibitem [{\citenamefont {Barnum}\ \emph {et~al.}(2014)\citenamefont {Barnum},
  \citenamefont {Müller},\ and\ \citenamefont {Ududec}}]{Barnum2014}%
  \BibitemOpen
  \bibfield  {author} {\bibinfo {author} {\bibfnamefont {Howard}\ \bibnamefont
  {Barnum}}, \bibinfo {author} {\bibfnamefont {Markus~P}\ \bibnamefont
  {Müller}}, \ and\ \bibinfo {author} {\bibfnamefont {Cozmin}\ \bibnamefont
  {Ududec}},\ }\bibfield  {title} {\enquote {\bibinfo {title} {Higher-order
  interference and single-system postulates characterizing quantum theory},}\
  }\href {\doibase 10.1088/1367-2630/16/12/123029} {\bibfield  {journal}
  {\bibinfo  {journal} {New Journal of Physics}\ }\textbf {\bibinfo {volume}
  {16}},\ \bibinfo {pages} {123029} (\bibinfo {year} {2014})}\BibitemShut
  {NoStop}%
\bibitem [{\citenamefont {Kauten}\ \emph {et~al.}(2017)\citenamefont {Kauten},
  \citenamefont {Keil}, \citenamefont {Kaufmann}, \citenamefont {Pressl},
  \citenamefont {Brukner},\ and\ \citenamefont {Weihs}}]{Kauten2017}%
  \BibitemOpen
  \bibfield  {author} {\bibinfo {author} {\bibfnamefont {Thomas}\ \bibnamefont
  {Kauten}}, \bibinfo {author} {\bibfnamefont {Robert}\ \bibnamefont {Keil}},
  \bibinfo {author} {\bibfnamefont {Thomas}\ \bibnamefont {Kaufmann}}, \bibinfo
  {author} {\bibfnamefont {Benedikt}\ \bibnamefont {Pressl}}, \bibinfo {author}
  {\bibfnamefont {{\v{C}}aslav}\ \bibnamefont {Brukner}}, \ and\ \bibinfo
  {author} {\bibfnamefont {Gregor}\ \bibnamefont {Weihs}},\ }\bibfield  {title}
  {\enquote {\bibinfo {title} {Obtaining tight bounds on higher-order
  interferences with a 5-path interferometer},}\ }\href {\doibase
  10.1088/1367-2630/aa5d98} {\bibfield  {journal} {\bibinfo  {journal} {New
  Journal of Physics}\ }\textbf {\bibinfo {volume} {19}},\ \bibinfo {pages}
  {033017} (\bibinfo {year} {2017})}\BibitemShut {NoStop}%
\bibitem [{\citenamefont {Feynman}\ \emph {et~al.}(1963)\citenamefont
  {Feynman}, \citenamefont {Leighton},\ and\ \citenamefont
  {Sands}}]{Feynman1963}%
  \BibitemOpen
  \bibfield  {author} {\bibinfo {author} {\bibfnamefont {R.~P.}\ \bibnamefont
  {Feynman}}, \bibinfo {author} {\bibfnamefont {R.}~\bibnamefont {Leighton}}, \
  and\ \bibinfo {author} {\bibfnamefont {M.}~\bibnamefont {Sands}},\
  }\href@noop {} {\emph {\bibinfo {title} {The Feynman Lectures on Physics}}},\
  Vol.~\bibinfo {volume} {3}\ (\bibinfo  {publisher} {Addison-Wesley, Reading
  Mass},\ \bibinfo {year} {1963})\BibitemShut {NoStop}%
\bibitem [{\citenamefont {Cohen-Tannoudji}\ \emph {et~al.}(2005)\citenamefont
  {Cohen-Tannoudji}, \citenamefont {Diu},\ and\ \citenamefont {Laloe}}]{Cohen}%
  \BibitemOpen
  \bibfield  {author} {\bibinfo {author} {\bibfnamefont {C.}~\bibnamefont
  {Cohen-Tannoudji}}, \bibinfo {author} {\bibfnamefont {B.}~\bibnamefont
  {Diu}}, \ and\ \bibinfo {author} {\bibfnamefont {F.}~\bibnamefont {Laloe}},\
  }\href@noop {} {\emph {\bibinfo {title} {Quantum Mechanics I, 2nd ed.}}},\
  Vol.~\bibinfo {volume} {1}\ (\bibinfo  {publisher} {Wiley-VCH},\ \bibinfo
  {year} {2005})\BibitemShut {NoStop}%
\bibitem [{\citenamefont {Shankar}(1994)}]{shankar1994}%
  \BibitemOpen
  \bibfield  {author} {\bibinfo {author} {\bibfnamefont {R}~\bibnamefont
  {Shankar}},\ }\bibfield  {title} {\enquote {\bibinfo {title} {Simple problems
  in one dimension},}\ }in\ \href@noop {} {\emph {\bibinfo {booktitle}
  {Principles of Quantum Mechanics}}}\ (\bibinfo  {publisher} {Springer},\
  \bibinfo {year} {1994})\ pp.\ \bibinfo {pages} {151--178}\BibitemShut
  {NoStop}%
\bibitem [{\citenamefont {J\"onsson}(1961)}]{Jonsson1961}%
  \BibitemOpen
  \bibfield  {author} {\bibinfo {author} {\bibfnamefont {Claus}\ \bibnamefont
  {J\"onsson}},\ }\bibfield  {title} {\enquote {\bibinfo {title}
  {Elektroneninterferenzen an mehreren künstlich hergestellten feinspalten},}\
  }\href {\doibase 10.1007/BF01342460} {\bibfield  {journal} {\bibinfo
  {journal} {Zeitschrift für Physik}\ }\textbf {\bibinfo {volume} {161}},\
  \bibinfo {pages} {454 -- 474} (\bibinfo {year} {1961})}\BibitemShut {NoStop}%
\bibitem [{\citenamefont {Zeilinger}\ \emph {et~al.}(1988)\citenamefont
  {Zeilinger}, \citenamefont {G\"ahler}, \citenamefont {Shull}, \citenamefont
  {Treimer},\ and\ \citenamefont {Mampe}}]{Zeilinger1988}%
  \BibitemOpen
  \bibfield  {author} {\bibinfo {author} {\bibfnamefont {Anton}\ \bibnamefont
  {Zeilinger}}, \bibinfo {author} {\bibfnamefont {Roland}\ \bibnamefont
  {G\"ahler}}, \bibinfo {author} {\bibfnamefont {C.~G.}\ \bibnamefont {Shull}},
  \bibinfo {author} {\bibfnamefont {Wolfgang}\ \bibnamefont {Treimer}}, \ and\
  \bibinfo {author} {\bibfnamefont {Walter}\ \bibnamefont {Mampe}},\ }\bibfield
   {title} {\enquote {\bibinfo {title} {Single- and double-slit diffraction of
  neutrons},}\ }\href {\doibase 10.1103/RevModPhys.60.1067} {\bibfield
  {journal} {\bibinfo  {journal} {Rev. Mod. Phys.}\ }\textbf {\bibinfo {volume}
  {60}},\ \bibinfo {pages} {1067--1073} (\bibinfo {year} {1988})}\BibitemShut
  {NoStop}%
\bibitem [{\citenamefont {Greenberger}\ and\ \citenamefont
  {Yasin}(1988)}]{greenberger1988}%
  \BibitemOpen
  \bibfield  {author} {\bibinfo {author} {\bibfnamefont {Daniel~M}\
  \bibnamefont {Greenberger}}\ and\ \bibinfo {author} {\bibfnamefont {Allaine}\
  \bibnamefont {Yasin}},\ }\bibfield  {title} {\enquote {\bibinfo {title}
  {Simultaneous wave and particle knowledge in a neutron interferometer},}\
  }\href@noop {} {\bibfield  {journal} {\bibinfo  {journal} {Physics Letters
  A}\ }\textbf {\bibinfo {volume} {128}},\ \bibinfo {pages} {391--394}
  (\bibinfo {year} {1988})}\BibitemShut {NoStop}%
\bibitem [{\citenamefont {Zeilinger}(1999)}]{Zeilinger1999}%
  \BibitemOpen
  \bibfield  {author} {\bibinfo {author} {\bibfnamefont {Anton}\ \bibnamefont
  {Zeilinger}},\ }\bibfield  {title} {\enquote {\bibinfo {title} {Experiment
  and the foundations of quantum physics},}\ }\href {\doibase
  10.1103/RevModPhys.71.S288} {\bibfield  {journal} {\bibinfo  {journal} {Rev.
  Mod. Phys.}\ }\textbf {\bibinfo {volume} {71}},\ \bibinfo {pages}
  {S288--S297} (\bibinfo {year} {1999})}\BibitemShut {NoStop}%
\bibitem [{\citenamefont {Feynman}\ and\ \citenamefont
  {Hibbs}(1965)}]{Feynman1965}%
  \BibitemOpen
  \bibfield  {author} {\bibinfo {author} {\bibfnamefont {R.~P.}\ \bibnamefont
  {Feynman}}\ and\ \bibinfo {author} {\bibfnamefont {A.~R.}\ \bibnamefont
  {Hibbs}},\ }\href@noop {} {\emph {\bibinfo {title} {Quantum Mechanics and
  Path Integrals, 3rd ed.}}}\ (\bibinfo  {publisher} {McGraw-Hill},\ \bibinfo
  {year} {1965})\BibitemShut {NoStop}%
\bibitem [{\citenamefont {Rengaraj}\ \emph {et~al.}(2018)\citenamefont
  {Rengaraj}, \citenamefont {Prathwiraj}, \citenamefont {Sahoo}, \citenamefont
  {Somashekhar},\ and\ \citenamefont {Sinha}}]{Rengaraj2018}%
  \BibitemOpen
  \bibfield  {author} {\bibinfo {author} {\bibfnamefont {G}~\bibnamefont
  {Rengaraj}}, \bibinfo {author} {\bibfnamefont {U}~\bibnamefont {Prathwiraj}},
  \bibinfo {author} {\bibfnamefont {Surya~Narayan}\ \bibnamefont {Sahoo}},
  \bibinfo {author} {\bibfnamefont {R}~\bibnamefont {Somashekhar}}, \ and\
  \bibinfo {author} {\bibfnamefont {Urbasi}\ \bibnamefont {Sinha}},\ }\bibfield
   {title} {\enquote {\bibinfo {title} {Measuring the deviation from the
  superposition principle in interference experiments},}\ }\href
  {http://stacks.iop.org/1367-2630/20/i=6/a=063049} {\bibfield  {journal}
  {\bibinfo  {journal} {New Journal of Physics}\ }\textbf {\bibinfo {volume}
  {20}},\ \bibinfo {pages} {063049} (\bibinfo {year} {2018})}\BibitemShut
  {NoStop}%
\bibitem [{\citenamefont {Taflove}\ and\ \citenamefont
  {Hagness}(2005)}]{Taflove}%
  \BibitemOpen
  \bibfield  {author} {\bibinfo {author} {\bibfnamefont {A.}~\bibnamefont
  {Taflove}}\ and\ \bibinfo {author} {\bibfnamefont {S.}~\bibnamefont
  {Hagness}},\ }\href@noop {} {\emph {\bibinfo {title} {Computational
  Electrodynamics: The Finite-Difference Time-Domain Method}}}\ (\bibinfo
  {publisher} {Artech House, Boston},\ \bibinfo {year} {2005})\BibitemShut
  {NoStop}%
\bibitem [{\citenamefont {De~Raedt}\ \emph {et~al.}(2012)\citenamefont
  {De~Raedt}, \citenamefont {Michielsen},\ and\ \citenamefont
  {Hess}}]{Raedt2012}%
  \BibitemOpen
  \bibfield  {author} {\bibinfo {author} {\bibfnamefont {Hans}\ \bibnamefont
  {De~Raedt}}, \bibinfo {author} {\bibfnamefont {Kristel}\ \bibnamefont
  {Michielsen}}, \ and\ \bibinfo {author} {\bibfnamefont {Karl}\ \bibnamefont
  {Hess}},\ }\bibfield  {title} {\enquote {\bibinfo {title} {Analysis of
  multipath interference in three-slit experiments},}\ }\href {\doibase
  10.1103/PhysRevA.85.012101} {\bibfield  {journal} {\bibinfo  {journal} {Phys.
  Rev. A}\ }\textbf {\bibinfo {volume} {85}},\ \bibinfo {pages} {012101}
  (\bibinfo {year} {2012})}\BibitemShut {NoStop}%
\bibitem [{\citenamefont {Sinha}\ \emph {et~al.}(2015)\citenamefont {Sinha},
  \citenamefont {Vijay},\ and\ \citenamefont {Sinha}}]{Sinha2015}%
  \BibitemOpen
  \bibfield  {author} {\bibinfo {author} {\bibfnamefont {Aninda}\ \bibnamefont
  {Sinha}}, \bibinfo {author} {\bibfnamefont {Aravind~H.}\ \bibnamefont
  {Vijay}}, \ and\ \bibinfo {author} {\bibfnamefont {Urbasi}\ \bibnamefont
  {Sinha}},\ }\bibfield  {title} {\enquote {\bibinfo {title} {On the
  superposition principle in interference experiments},}\ }\href {\doibase
  10.1038/srep10304} {\bibfield  {journal} {\bibinfo  {journal} {Scientific
  Reports}\ }\textbf {\bibinfo {volume} {5}},\ \bibinfo {pages} {10304}
  (\bibinfo {year} {2015})}\BibitemShut {NoStop}%
\bibitem [{\citenamefont {Wootters}\ and\ \citenamefont
  {Zurek}(1979)}]{Wooters1979}%
  \BibitemOpen
  \bibfield  {author} {\bibinfo {author} {\bibfnamefont {William~K.}\
  \bibnamefont {Wootters}}\ and\ \bibinfo {author} {\bibfnamefont
  {Wojciech~H.}\ \bibnamefont {Zurek}},\ }\bibfield  {title} {\enquote
  {\bibinfo {title} {Complementarity in the double-slit experiment: Quantum
  nonseparability and a quantitative statement of bohr's principle},}\ }\href
  {\doibase 10.1103/PhysRevD.19.473} {\bibfield  {journal} {\bibinfo  {journal}
  {Phys. Rev. D}\ }\textbf {\bibinfo {volume} {19}},\ \bibinfo {pages}
  {473--484} (\bibinfo {year} {1979})}\BibitemShut {NoStop}%
\bibitem [{\citenamefont {Scully}\ \emph {et~al.}(1991)\citenamefont {Scully},
  \citenamefont {Englert},\ and\ \citenamefont {Walther}}]{Scully1991}%
  \BibitemOpen
  \bibfield  {author} {\bibinfo {author} {\bibfnamefont {Marian~O.}\
  \bibnamefont {Scully}}, \bibinfo {author} {\bibfnamefont {Berthold-Georg}\
  \bibnamefont {Englert}}, \ and\ \bibinfo {author} {\bibfnamefont {Herbert}\
  \bibnamefont {Walther}},\ }\bibfield  {title} {\enquote {\bibinfo {title}
  {Quantum optical tests of complementarity},}\ }\href {\doibase
  http://dx.doi.org/10.1038/351111a0} {\bibfield  {journal} {\bibinfo
  {journal} {Nature}\ }\textbf {\bibinfo {volume} {351}},\ \bibinfo {pages}
  {111--116} (\bibinfo {year} {1991})}\BibitemShut {NoStop}%
\bibitem [{\citenamefont {Scully}\ and\ \citenamefont
  {Dr\"uhl}(1982)}]{Scully1982}%
  \BibitemOpen
  \bibfield  {author} {\bibinfo {author} {\bibfnamefont {Marlan~O.}\
  \bibnamefont {Scully}}\ and\ \bibinfo {author} {\bibfnamefont {Kai}\
  \bibnamefont {Dr\"uhl}},\ }\bibfield  {title} {\enquote {\bibinfo {title}
  {Quantum eraser: A proposed photon correlation experiment concerning
  observation and "delayed choice" in quantum mechanics},}\ }\href {\doibase
  10.1103/PhysRevA.25.2208} {\bibfield  {journal} {\bibinfo  {journal} {Phys.
  Rev. A}\ }\textbf {\bibinfo {volume} {25}},\ \bibinfo {pages} {2208--2213}
  (\bibinfo {year} {1982})}\BibitemShut {NoStop}%
\bibitem [{\citenamefont {Wheeler}(1983)}]{Wheeler}%
  \BibitemOpen
  \bibfield  {author} {\bibinfo {author} {\bibfnamefont {J.A.}\ \bibnamefont
  {Wheeler}},\ }\href@noop {} {\emph {\bibinfo {title} {Quantum Theory and
  Measurement}}}\ (\bibinfo  {publisher} {Princeton University Press},\
  \bibinfo {year} {1983})\BibitemShut {NoStop}%
\bibitem [{\citenamefont {Herzog}\ \emph {et~al.}(1995)\citenamefont {Herzog},
  \citenamefont {Kwiat}, \citenamefont {Weinfurter},\ and\ \citenamefont
  {Zeilinger}}]{Zeilinger}%
  \BibitemOpen
  \bibfield  {author} {\bibinfo {author} {\bibfnamefont {Thomas~J.}\
  \bibnamefont {Herzog}}, \bibinfo {author} {\bibfnamefont {Paul~G.}\
  \bibnamefont {Kwiat}}, \bibinfo {author} {\bibfnamefont {Harald}\
  \bibnamefont {Weinfurter}}, \ and\ \bibinfo {author} {\bibfnamefont {Anton}\
  \bibnamefont {Zeilinger}},\ }\bibfield  {title} {\enquote {\bibinfo {title}
  {Complementarity and the quantum eraser},}\ }\href {\doibase
  10.1103/PhysRevLett.75.3034} {\bibfield  {journal} {\bibinfo  {journal}
  {Phys. Rev. Lett.}\ }\textbf {\bibinfo {volume} {75}},\ \bibinfo {pages}
  {3034--3037} (\bibinfo {year} {1995})}\BibitemShut {NoStop}%
\bibitem [{\citenamefont {Kim}\ \emph {et~al.}(2000)\citenamefont {Kim},
  \citenamefont {Yu}, \citenamefont {Kulik}, \citenamefont {Shih},\ and\
  \citenamefont {Scully}}]{Kim}%
  \BibitemOpen
  \bibfield  {author} {\bibinfo {author} {\bibfnamefont {Yoon-Ho}\ \bibnamefont
  {Kim}}, \bibinfo {author} {\bibfnamefont {Rong}\ \bibnamefont {Yu}}, \bibinfo
  {author} {\bibfnamefont {Sergei~P.}\ \bibnamefont {Kulik}}, \bibinfo {author}
  {\bibfnamefont {Yanhua}\ \bibnamefont {Shih}}, \ and\ \bibinfo {author}
  {\bibfnamefont {Marlan~O.}\ \bibnamefont {Scully}},\ }\bibfield  {title}
  {\enquote {\bibinfo {title} {Delayed ``choice'' quantum eraser},}\ }\href
  {\doibase 10.1103/PhysRevLett.84.1} {\bibfield  {journal} {\bibinfo
  {journal} {Phys. Rev. Lett.}\ }\textbf {\bibinfo {volume} {84}},\ \bibinfo
  {pages} {1--5} (\bibinfo {year} {2000})}\BibitemShut {NoStop}%
\bibitem [{\citenamefont {Jacques}\ \emph {et~al.}(2007)\citenamefont
  {Jacques}, \citenamefont {Wu}, \citenamefont {Grosshans}, \citenamefont
  {Treussart}, \citenamefont {Grangier}, \citenamefont {Aspect},\ and\
  \citenamefont {Roch}}]{Wu}%
  \BibitemOpen
  \bibfield  {author} {\bibinfo {author} {\bibfnamefont {Vincent}\ \bibnamefont
  {Jacques}}, \bibinfo {author} {\bibfnamefont {E}~\bibnamefont {Wu}}, \bibinfo
  {author} {\bibfnamefont {Fr{\'e}d{\'e}ric}\ \bibnamefont {Grosshans}},
  \bibinfo {author} {\bibfnamefont {Fran{\c c}ois}\ \bibnamefont {Treussart}},
  \bibinfo {author} {\bibfnamefont {Philippe}\ \bibnamefont {Grangier}},
  \bibinfo {author} {\bibfnamefont {Alain}\ \bibnamefont {Aspect}}, \ and\
  \bibinfo {author} {\bibfnamefont {Jean-Fran{\c c}ois}\ \bibnamefont {Roch}},\
  }\bibfield  {title} {\enquote {\bibinfo {title} {Experimental realization of
  wheeler{\textquoteright}s delayed-choice gedanken experiment},}\ }\href
  {\doibase 10.1126/science.1136303} {\bibfield  {journal} {\bibinfo  {journal}
  {Science}\ }\textbf {\bibinfo {volume} {315}},\ \bibinfo {pages} {966--968}
  (\bibinfo {year} {2007})},\ \Eprint
  {http://arxiv.org/abs/https://science.sciencemag.org/content/315/5814/966.full.pdf}
  {https://science.sciencemag.org/content/315/5814/966.full.pdf} \BibitemShut
  {NoStop}%
\bibitem [{\citenamefont {Walborn}\ \emph {et~al.}(2002)\citenamefont
  {Walborn}, \citenamefont {Terra~Cunha}, \citenamefont {P\'adua},\ and\
  \citenamefont {Monken}}]{Walborn2002}%
  \BibitemOpen
  \bibfield  {author} {\bibinfo {author} {\bibfnamefont {S.~P.}\ \bibnamefont
  {Walborn}}, \bibinfo {author} {\bibfnamefont {M.~O.}\ \bibnamefont
  {Terra~Cunha}}, \bibinfo {author} {\bibfnamefont {S.}~\bibnamefont
  {P\'adua}}, \ and\ \bibinfo {author} {\bibfnamefont {C.~H.}\ \bibnamefont
  {Monken}},\ }\bibfield  {title} {\enquote {\bibinfo {title} {Double-slit
  quantum eraser},}\ }\href {\doibase 10.1103/PhysRevA.65.033818} {\bibfield
  {journal} {\bibinfo  {journal} {Phys. Rev. A}\ }\textbf {\bibinfo {volume}
  {65}},\ \bibinfo {pages} {033818} (\bibinfo {year} {2002})}\BibitemShut
  {NoStop}%
\bibitem [{\citenamefont {Pavičić}(1996)}]{Pavicic1996}%
  \BibitemOpen
  \bibfield  {author} {\bibinfo {author} {\bibfnamefont {Mladen}\ \bibnamefont
  {Pavičić}},\ }\bibfield  {title} {\enquote {\bibinfo {title} {Resonance
  energy-exchange-free detection and “welcher weg” experiment},}\ }\href
  {\doibase https://doi.org/10.1016/S0375-9601(96)00751-7} {\bibfield
  {journal} {\bibinfo  {journal} {Physics Letters A}\ }\textbf {\bibinfo
  {volume} {223}},\ \bibinfo {pages} {241 -- 245} (\bibinfo {year}
  {1996})}\BibitemShut {NoStop}%
\bibitem [{\citenamefont {de~Oliveira}\ \emph {et~al.}(2017)\citenamefont
  {de~Oliveira}, \citenamefont {de~Souza}, \citenamefont {Cabral},
  \citenamefont {da~Paz},\ and\ \citenamefont {Sampaio}}]{deOliveira2017}%
  \BibitemOpen
  \bibfield  {author} {\bibinfo {author} {\bibfnamefont {J.G.G.}\ \bibnamefont
  {de~Oliveira}}, \bibinfo {author} {\bibfnamefont {Gustavo}\ \bibnamefont
  {de~Souza}}, \bibinfo {author} {\bibfnamefont {L.A.}\ \bibnamefont {Cabral}},
  \bibinfo {author} {\bibfnamefont {I.G.}\ \bibnamefont {da~Paz}}, \ and\
  \bibinfo {author} {\bibfnamefont {Marcos}\ \bibnamefont {Sampaio}},\
  }\bibfield  {title} {\enquote {\bibinfo {title} {Exotic looped trajectories
  via quantum marking},}\ }\href {\doibase
  https://doi.org/10.1016/j.aop.2017.10.006} {\bibfield  {journal} {\bibinfo
  {journal} {Annals of Physics}\ }\textbf {\bibinfo {volume} {387}},\ \bibinfo
  {pages} {222 -- 238} (\bibinfo {year} {2017})}\BibitemShut {NoStop}%
\bibitem [{\citenamefont {Born}(1926)}]{Born1926}%
  \BibitemOpen
  \bibfield  {author} {\bibinfo {author} {\bibfnamefont {Max}\ \bibnamefont
  {Born}},\ }\bibfield  {title} {\enquote {\bibinfo {title} {Quantenmechanik
  der sto{\ss}vorg{\"a}nge},}\ }\href {\doibase 10.1007/BF01397184} {\bibfield
  {journal} {\bibinfo  {journal} {Zeitschrift f{\"u}r Physik}\ }\textbf
  {\bibinfo {volume} {38}},\ \bibinfo {pages} {803--827} (\bibinfo {year}
  {1926})}\BibitemShut {NoStop}%
\bibitem [{\citenamefont {Sinha}\ \emph {et~al.}(2010)\citenamefont {Sinha},
  \citenamefont {Couteau}, \citenamefont {Jennewein}, \citenamefont
  {Laflamme},\ and\ \citenamefont {Weihs}}]{Sinha2010}%
  \BibitemOpen
  \bibfield  {author} {\bibinfo {author} {\bibfnamefont {Urbasi}\ \bibnamefont
  {Sinha}}, \bibinfo {author} {\bibfnamefont {Christophe}\ \bibnamefont
  {Couteau}}, \bibinfo {author} {\bibfnamefont {Thomas}\ \bibnamefont
  {Jennewein}}, \bibinfo {author} {\bibfnamefont {Raymond}\ \bibnamefont
  {Laflamme}}, \ and\ \bibinfo {author} {\bibfnamefont {Gregor}\ \bibnamefont
  {Weihs}},\ }\bibfield  {title} {\enquote {\bibinfo {title} {Ruling out
  multi-order interference in quantum mechanics},}\ }\href {\doibase
  10.1126/science.1190545} {\bibfield  {journal} {\bibinfo  {journal}
  {Science}\ }\textbf {\bibinfo {volume} {329}},\ \bibinfo {pages} {418--421}
  (\bibinfo {year} {2010})}\BibitemShut {NoStop}%
\bibitem [{\citenamefont {Sorkin}(1994)}]{Sorkin1994}%
  \BibitemOpen
  \bibfield  {author} {\bibinfo {author} {\bibfnamefont {Rafael~D.}\
  \bibnamefont {Sorkin}},\ }\bibfield  {title} {\enquote {\bibinfo {title}
  {Quantum mechanics as a quantum measure theory},}\ }\href {\doibase
  https://doi.org/10.1142/S021773239400294X} {\bibfield  {journal} {\bibinfo
  {journal} {Mod. Phys. Lett.}\ }\textbf {\bibinfo {volume} {A9}},\ \bibinfo
  {pages} {3119--3128} (\bibinfo {year} {1994})}\BibitemShut {NoStop}%
\bibitem [{\citenamefont {Maga\~{n}a Loaiza}\ \emph {et~al.}(2016)\citenamefont
  {Maga\~{n}a Loaiza}, \citenamefont {De~Leon}, \citenamefont {Mirhosseini},
  \citenamefont {Fickler}, \citenamefont {Safari}, \citenamefont {Mick},
  \citenamefont {McIntyre}, \citenamefont {Banzer}, \citenamefont {Rodenburg},
  \citenamefont {Leuchs},\ and\ \citenamefont {Boyd}}]{Magana2016}%
  \BibitemOpen
  \bibfield  {author} {\bibinfo {author} {\bibfnamefont {O.~S.}\ \bibnamefont
  {Maga\~{n}a Loaiza}}, \bibinfo {author} {\bibfnamefont {I.}~\bibnamefont
  {De~Leon}}, \bibinfo {author} {\bibfnamefont {M.}~\bibnamefont
  {Mirhosseini}}, \bibinfo {author} {\bibfnamefont {R.}~\bibnamefont
  {Fickler}}, \bibinfo {author} {\bibfnamefont {A.}~\bibnamefont {Safari}},
  \bibinfo {author} {\bibfnamefont {U.}~\bibnamefont {Mick}}, \bibinfo {author}
  {\bibfnamefont {B.}~\bibnamefont {McIntyre}}, \bibinfo {author}
  {\bibfnamefont {P.}~\bibnamefont {Banzer}}, \bibinfo {author} {\bibfnamefont
  {B.}~\bibnamefont {Rodenburg}}, \bibinfo {author} {\bibfnamefont
  {G.}~\bibnamefont {Leuchs}}, \ and\ \bibinfo {author} {\bibfnamefont {R.~W.}\
  \bibnamefont {Boyd}},\ }\bibfield  {title} {\enquote {\bibinfo {title}
  {Exotic looped trajectories of photons in three-slit interference},}\ }\href
  {\doibase 10.1142/S021773231950233X} {\bibfield  {journal} {\bibinfo
  {journal} {Nature Communications}\ }\textbf {\bibinfo {volume} {7}},\
  \bibinfo {pages} {13987} (\bibinfo {year} {2016})}\BibitemShut {NoStop}%
\bibitem [{\citenamefont {Sawant}\ \emph {et~al.}(2014)\citenamefont {Sawant},
  \citenamefont {Samuel}, \citenamefont {Sinha}, \citenamefont {Sinha},\ and\
  \citenamefont {Sinha}}]{Sawant2014}%
  \BibitemOpen
  \bibfield  {author} {\bibinfo {author} {\bibfnamefont {Rahul}\ \bibnamefont
  {Sawant}}, \bibinfo {author} {\bibfnamefont {Joseph}\ \bibnamefont {Samuel}},
  \bibinfo {author} {\bibfnamefont {Aninda}\ \bibnamefont {Sinha}}, \bibinfo
  {author} {\bibfnamefont {Supurna}\ \bibnamefont {Sinha}}, \ and\ \bibinfo
  {author} {\bibfnamefont {Urbasi}\ \bibnamefont {Sinha}},\ }\bibfield  {title}
  {\enquote {\bibinfo {title} {Nonclassical paths in quantum interference
  experiments},}\ }\href {\doibase 10.1103/PhysRevLett.113.120406} {\bibfield
  {journal} {\bibinfo  {journal} {Phys. Rev. Lett.}\ }\textbf {\bibinfo
  {volume} {113}},\ \bibinfo {pages} {120406} (\bibinfo {year}
  {2014})}\BibitemShut {NoStop}%
\bibitem [{\citenamefont {Quach}(2017)}]{Quach2017}%
  \BibitemOpen
  \bibfield  {author} {\bibinfo {author} {\bibfnamefont {James~Q.}\
  \bibnamefont {Quach}},\ }\bibfield  {title} {\enquote {\bibinfo {title}
  {Which-way double-slit experiments and born-rule violation},}\ }\href
  {\doibase 10.1103/PhysRevA.95.042129} {\bibfield  {journal} {\bibinfo
  {journal} {Phys. Rev. A}\ }\textbf {\bibinfo {volume} {95}},\ \bibinfo
  {pages} {042129} (\bibinfo {year} {2017})}\BibitemShut {NoStop}%
\bibitem [{\citenamefont {da~Paz}\ \emph {et~al.}(2016)\citenamefont {da~Paz},
  \citenamefont {Vieira}, \citenamefont {Ducharme}, \citenamefont {Cabral},
  \citenamefont {Alexander},\ and\ \citenamefont {Sampaio}}]{daPaz2016}%
  \BibitemOpen
  \bibfield  {author} {\bibinfo {author} {\bibfnamefont {I.~G.}\ \bibnamefont
  {da~Paz}}, \bibinfo {author} {\bibfnamefont {C.~H.~S.}\ \bibnamefont
  {Vieira}}, \bibinfo {author} {\bibfnamefont {R.}~\bibnamefont {Ducharme}},
  \bibinfo {author} {\bibfnamefont {L.~A.}\ \bibnamefont {Cabral}}, \bibinfo
  {author} {\bibfnamefont {H.}~\bibnamefont {Alexander}}, \ and\ \bibinfo
  {author} {\bibfnamefont {M.~D.~R.}\ \bibnamefont {Sampaio}},\ }\bibfield
  {title} {\enquote {\bibinfo {title} {Gouy phase in nonclassical paths in a
  triple-slit interference experiment},}\ }\href {\doibase
  10.1103/PhysRevA.93.033621} {\bibfield  {journal} {\bibinfo  {journal} {Phys.
  Rev. A}\ }\textbf {\bibinfo {volume} {93}},\ \bibinfo {pages} {033621}
  (\bibinfo {year} {2016})}\BibitemShut {NoStop}%
\bibitem [{\citenamefont {Vieira}\ \emph {et~al.}(2019)\citenamefont {Vieira},
  \citenamefont {Costa}, \citenamefont {de~Souza}, \citenamefont {Sampaio},\
  and\ \citenamefont {da~Paz}}]{Vieira2019}%
  \BibitemOpen
  \bibfield  {author} {\bibinfo {author} {\bibfnamefont {C.~H.~S.}\
  \bibnamefont {Vieira}}, \bibinfo {author} {\bibfnamefont {H.~A.~S.}\
  \bibnamefont {Costa}}, \bibinfo {author} {\bibfnamefont {G.}~\bibnamefont
  {de~Souza}}, \bibinfo {author} {\bibfnamefont {M.}~\bibnamefont {Sampaio}}, \
  and\ \bibinfo {author} {\bibfnamefont {I.~G.}\ \bibnamefont {da~Paz}},\
  }\bibfield  {title} {\enquote {\bibinfo {title} {Fringe visibility of exotic
  trajectories for matter waves in a double-slit experiment},}\ }\href
  {\doibase 10.1142/S021773231950233X} {\bibfield  {journal} {\bibinfo
  {journal} {Modern Physics Letters A}\ }\textbf {\bibinfo {volume} {34}},\
  \bibinfo {pages} {1950233} (\bibinfo {year} {2019})},\ \Eprint
  {http://arxiv.org/abs/https://doi.org/10.1142/S021773231950233X}
  {https://doi.org/10.1142/S021773231950233X} \BibitemShut {NoStop}%
\bibitem [{\citenamefont {Krause}\ \emph {et~al.}(1987)\citenamefont {Krause},
  \citenamefont {Scully},\ and\ \citenamefont {Walther}}]{Krause87}%
  \BibitemOpen
  \bibfield  {author} {\bibinfo {author} {\bibfnamefont {Joachim}\ \bibnamefont
  {Krause}}, \bibinfo {author} {\bibfnamefont {Marlan~O.}\ \bibnamefont
  {Scully}}, \ and\ \bibinfo {author} {\bibfnamefont {Herbert}\ \bibnamefont
  {Walther}},\ }\bibfield  {title} {\enquote {\bibinfo {title} {State reduction
  and |n> - state preparation in a high - q micromaser},}\ }\href {\doibase
  10.1103/PhysRevA.36.4547} {\bibfield  {journal} {\bibinfo  {journal} {Phys.
  Rev. A}\ }\textbf {\bibinfo {volume} {36}},\ \bibinfo {pages} {4547--4550}
  (\bibinfo {year} {1987})}\BibitemShut {NoStop}%
\bibitem [{\citenamefont {Srinivas}\ and\ \citenamefont
  {Davies}(1981)}]{Srinivas1981}%
  \BibitemOpen
  \bibfield  {author} {\bibinfo {author} {\bibfnamefont {M.D.}\ \bibnamefont
  {Srinivas}}\ and\ \bibinfo {author} {\bibfnamefont {E.B.}\ \bibnamefont
  {Davies}},\ }\bibfield  {title} {\enquote {\bibinfo {title} {Photon counting
  probabilities in quantum optics},}\ }\href {\doibase 10.1080/713820643}
  {\bibfield  {journal} {\bibinfo  {journal} {Optica Acta: International
  Journal of Optics}\ }\textbf {\bibinfo {volume} {28}},\ \bibinfo {pages}
  {981--996} (\bibinfo {year} {1981})},\ \Eprint
  {http://arxiv.org/abs/https://doi.org/10.1080/713820643}
  {https://doi.org/10.1080/713820643} \BibitemShut {NoStop}%
\bibitem [{\citenamefont {Zoller}\ \emph {et~al.}(1987)\citenamefont {Zoller},
  \citenamefont {Marte},\ and\ \citenamefont {Walls}}]{Zoller1987}%
  \BibitemOpen
  \bibfield  {author} {\bibinfo {author} {\bibfnamefont {P.}~\bibnamefont
  {Zoller}}, \bibinfo {author} {\bibfnamefont {M.}~\bibnamefont {Marte}}, \
  and\ \bibinfo {author} {\bibfnamefont {D.~F.}\ \bibnamefont {Walls}},\
  }\bibfield  {title} {\enquote {\bibinfo {title} {Quantum jumps in atomic
  systems},}\ }\href {\doibase 10.1103/PhysRevA.35.198} {\bibfield  {journal}
  {\bibinfo  {journal} {Phys. Rev. A}\ }\textbf {\bibinfo {volume} {35}},\
  \bibinfo {pages} {198--207} (\bibinfo {year} {1987})}\BibitemShut {NoStop}%
\bibitem [{\citenamefont {Aharonov}\ and\ \citenamefont
  {Bohm}(1959)}]{aharonov1959}%
  \BibitemOpen
  \bibfield  {author} {\bibinfo {author} {\bibfnamefont {Yakir}\ \bibnamefont
  {Aharonov}}\ and\ \bibinfo {author} {\bibfnamefont {David}\ \bibnamefont
  {Bohm}},\ }\bibfield  {title} {\enquote {\bibinfo {title} {Significance of
  electromagnetic potentials in the quantum theory},}\ }\href@noop {}
  {\bibfield  {journal} {\bibinfo  {journal} {Physical Review}\ }\textbf
  {\bibinfo {volume} {115}},\ \bibinfo {pages} {485} (\bibinfo {year}
  {1959})}\BibitemShut {NoStop}%
\bibitem [{\citenamefont {Assafr\~{a}o}\ \emph {et~al.}(2019)\citenamefont
  {Assafr\~{a}o}, \citenamefont {Favarato}, \citenamefont {Gonçalves},\ and\
  \citenamefont {Simonelli}}]{Assafrao2019}%
  \BibitemOpen
  \bibfield  {author} {\bibinfo {author} {\bibfnamefont {D.}~\bibnamefont
  {Assafr\~{a}o}}, \bibinfo {author} {\bibfnamefont {C.~C.}\ \bibnamefont
  {Favarato}}, \bibinfo {author} {\bibfnamefont {S.~V.~B.}\ \bibnamefont
  {Gonçalves}}, \ and\ \bibinfo {author} {\bibfnamefont {G.}~\bibnamefont
  {Simonelli}},\ }\bibfield  {title} {\enquote {\bibinfo {title} {The
  double-slit electron diffraction experiment with aharonov-bohm phase effect
  revisited and the divergence in its asymptotic form},}\ }\href@noop {}
  {\bibfield  {journal} {\bibinfo  {journal} {Brazilian Journal of Physics}\
  }\textbf {\bibinfo {volume} {49}},\ \bibinfo {pages} {301--313} (\bibinfo
  {year} {2019})}\BibitemShut {NoStop}%
\bibitem [{\citenamefont {Davies}(1976)}]{Davies1976}%
  \BibitemOpen
  \bibfield  {author} {\bibinfo {author} {\bibfnamefont {E.~B.}\ \bibnamefont
  {Davies}},\ }\href@noop {} {\emph {\bibinfo {title} {Quantum Theory of Open
  Systems}}}\ (\bibinfo  {publisher} {Academic Press},\ \bibinfo {year}
  {1976})\BibitemShut {NoStop}%
\bibitem [{\citenamefont {Jacobs}(2010)}]{Jacobs2010}%
  \BibitemOpen
  \bibfield  {author} {\bibinfo {author} {\bibfnamefont {Kurt}\ \bibnamefont
  {Jacobs}},\ }\href@noop {} {\emph {\bibinfo {title} {Stochastic Process for
  Physicists - Understanding noisy systems}}}\ (\bibinfo  {publisher}
  {University Press},\ \bibinfo {year} {2010})\BibitemShut {NoStop}%
\end{thebibliography}%

\newpage
\end{document}